%% file: main.tex
\documentclass[10pt,conference]{IEEEtran}
\usepackage{cite}
\usepackage{amsmath,amssymb,amsfonts}
\usepackage{algorithmic}
\usepackage{graphicx}
\usepackage{textcomp}
\usepackage{xcolor}
\usepackage[hyphens]{url}
\usepackage{fancyhdr}
\usepackage{hyperref}

\usepackage{xspace}
\usepackage{multirow}
\usepackage[ruled,boxed,linesnumbered,noend]{algorithm2e}
\usepackage{enumitem}
\usepackage{xfrac}
\usepackage{booktabs}
\usepackage[table]{xcolor}
\usepackage{tabularx}
\usepackage{changepage}  
\usepackage{multicol}

\newtheorem{observation}{Observation}

\newcommand{\fcRone}[1]{\textcolor{black}{#1}}
\newcommand{\fcRtwo}[1]{\textcolor{black}{#1}}
\newcommand{\fcRthree}[1]{\textcolor{black}{#1}}

\newcommand{\hpcayear}{2027}

\newcommand{\projName}{Glimpse\xspace}

\newcommand{\hpcasubmissionnumber}{1803}
\title{\projName: Exploiting Draft Attention for Bandwidth-Efficient Verification in Mobile Edge Speculative Decoding}

\def\hpcacameraready{} 
\newcommand\hpcaauthors{Chao Fang$^{\dagger,*}$, Xiaodong Liu$^{\ddagger}$, Robin Geens$^\dagger$, Xiaoling Yi$^\dagger$, Arne Symons$^\dagger$, Yuqi Jiang$^{\ddagger}$, Marian Verhelst$^\dagger$}
\newcommand\hpcaaffiliation{$^\dagger$ESAT-MICAS, KU Leuven, Belgium\\$^\ddagger$IC Department, vivo Mobile Communication Co., Ltd, China}
\newcommand\hpcaemail{Email: \{chao.fang, robin.geens, xiaoling.yi, arne.symons, marian.verhelst\}@kuleuven.be, \{damon, jiangyuqi\}@vivo.com}

\input{hpca-template}

\begin{document}


\maketitle




\renewcommand{\thefootnote}{}
\footnotetext{$^*$Corresponding author}

\input{chapters/0-abstract}
\input{chapters/1-introduction}
\input{chapters/2-motivation}
\input{chapters/3-algorithm}

\input{chapters/4-hardware}
\input{chapters/5-evaluation}
\input{chapters/6-related}
\input{chapters/7-conclusion}

\section*{Acknowledgements}
This project has been partly funded by the European Research Council (ERC) under grant agreement No. 101088865, the Flanders AI Research Program, and long-term structural Methusalem funding by the Flemish Government. 
The authors acknowledge computing resources provided by the NVIDIA Academic Grant Program, by the EuroHPC Joint Undertaking through a Development Access call on the EuroHPC supercomputer LEONARDO, hosted by CINECA (Italy) and the LEONARDO consortium, and by the VSC (Flemish Supercomputer Center), funded by the Research Foundation - Flanders (FWO) and the Flemish Government.


\bibliographystyle{IEEEtranS}
\bibliography{refs}
\input{chapters/10-AI_use}

\end{document}

%% file: hpca-template.tex
\author{
  \ifdefined\hpcacameraready
    \IEEEauthorblockN{\hpcaauthors{}}
      \IEEEauthorblockA{
        \hpcaaffiliation{} \\
        \hpcaemail{}
      }
  \else
    \IEEEauthorblockN{\normalsize{HPCA \hpcayear{} Submission
      \textbf{\#\hpcasubmissionnumber{}}} \\
      \IEEEauthorblockA{
        Confidential Draft \\
        Do NOT Distribute!!
      }
    }
  \fi 
}

\fancypagestyle{camerareadyfirstpage}{%
  \fancyhead{}
  
  \fancyhead[C]{
    \ifdefined\aeopen
    \parbox[][12mm][t]{13.5cm}{\hpcayear{} IEEE International Symposium on High-Performance Computer Architecture (HPCA)}    
    \else
      \ifdefined\aereviewed
      \parbox[][12mm][t]{13.5cm}{\hpcayear{} IEEE International Symposium on High-Performance Computer Architecture (HPCA)}
      \else
      \ifdefined\aereproduced
      \parbox[][12mm][t]{13.5cm}{\hpcayear{} IEEE International Symposium on High-Performance Computer Architecture (HPCA)}
      \else
      \parbox[][0mm][t]{13.5cm}{\hpcayear{} IEEE International Symposium on High-Performance Computer Architecture (HPCA)}
    \fi 
    \fi 
    \fi 
    \ifdefined\aeopen 
      \includegraphics[width=12mm,height=12mm]{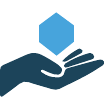}
    \fi 
    \ifdefined\aereviewed
      \includegraphics[width=12mm,height=12mm]{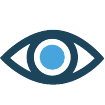}
    \fi 
    \ifdefined\aereproduced
      \includegraphics[width=12mm,height=12mm]{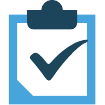}
    \fi
  }
  \fancyfoot[C]{}
}
\ifdefined\hpcacameraready 
\else
\fi

\ifdefined\eaopen

\fi

%% file: chapters/0-abstract.tex
\begin{abstract}
\fcRone{Large language models (LLMs) deployed on edge devices face a severe memory-bandwidth bottleneck during autoregressive decoding, which is further exacerbated by the ever-growing KV cache of emerging agentic workloads.}
\fcRone{Tree-based speculative decoding (SD) only shifts this bottleneck to verification, where the full prefix KV cache must be reloaded at every step.}
\fcRone{We observe that most prefix tokens contribute negligibly to verification yet cost the same to load, while the draft model's attention scores offer a free importance proxy for selective KV loading.}
\fcRone{Building on this insight, we present \projName, an algorithm-hardware co-design that turns draft attention into verification bandwidth savings for \fcRtwo{mobile edge} speculative decoding.}
\fcRone{At the algorithm level, \projName introduces a two-pass verification scheme.}
\fcRone{A lossy pass loads only the top-ranked prefix chunks guided by accumulated draft scores, while a periodic lossless pass merges the lossy-accepted chain with a reduced draft tree into one full-attention re-verification, \fcRthree{restoring the target model's greedy output} and amortizing its bandwidth cost over many accepted tokens.}
\fcRone{At the hardware level, \projName extends a conventional edge NPU with a lightweight Importance Selection Unit that performs on-chip score accumulation and top-K chunk selection at negligible overhead.}
\fcRone{Evaluated on six target LLMs and three benchmarks against the state-of-the-art (SotA) EAGLE-3, \projName delivers \fcRtwo{1.33--1.74$\times$} speedup and \fcRtwo{1.40--1.76$\times$} energy efficiency \fcRtwo{across 2K--64K contexts}, with only 2.7\% area overhead over the baseline edge NPU.}

\end{abstract}

%% file: chapters/1-introduction.tex
\section{Introduction} \label{sec:introduction}

Large language models (LLMs) are evolving from passive question-answering tools into on-device agents that plan autonomously, invoke tools, and reason over extended interactions~\cite{AppleIntelligence, SamsungGalaxyAI, AppAgent, MobileAgent}.
\fcRone{These agents routinely access chat histories, screen content, and local files, all of which involve highly sensitive personal data, making on-device inference essential for both latency~\cite{MobileSoC, EdgeLLM, FastTTS} and privacy~\cite{PrivacyLLM, AgentDAM, AppleIntelligence}.}
\fcRone{A key characteristic of agentic workloads is that context accumulates continuously across turns and tool calls}\fcRtwo{, and long reasoning traces extend it further within each turn}\fcRone{, often exceeding tens of thousands of tokens within a single session~\cite{LongSpec, MobileSoC, FastTTS, EdgeReasoning}.}
\fcRone{To run these long-context workloads on edge devices, weights are typically quantized to low bit-width (e.g., 4-bit)~\cite{GPTQ, AWQ, OmniQuant, Anda, FIGNA, BitMoD, COMET}, while the KV cache is kept at higher precision (e.g., 16-bit) to preserve generation quality.}
\fcRtwo{As shown in Table~\ref{tab:hw-spec}, mobile edge platforms offer only 51--120\,GB/s of DRAM bandwidth, one to two orders of magnitude below cloud GPUs. Loading the ever-growing KV cache thus increasingly dominates decode latency and energy on these devices~\cite{EdgeLLM}.}
\fcRtwo{Moreover, these platforms operate on batteries of merely 12.7--53.8\,Wh within power budgets around 11--40\,W, making per-token energy as critical as latency.}

\begin{figure}[t!]
    \centering
    \includegraphics[width=\linewidth]{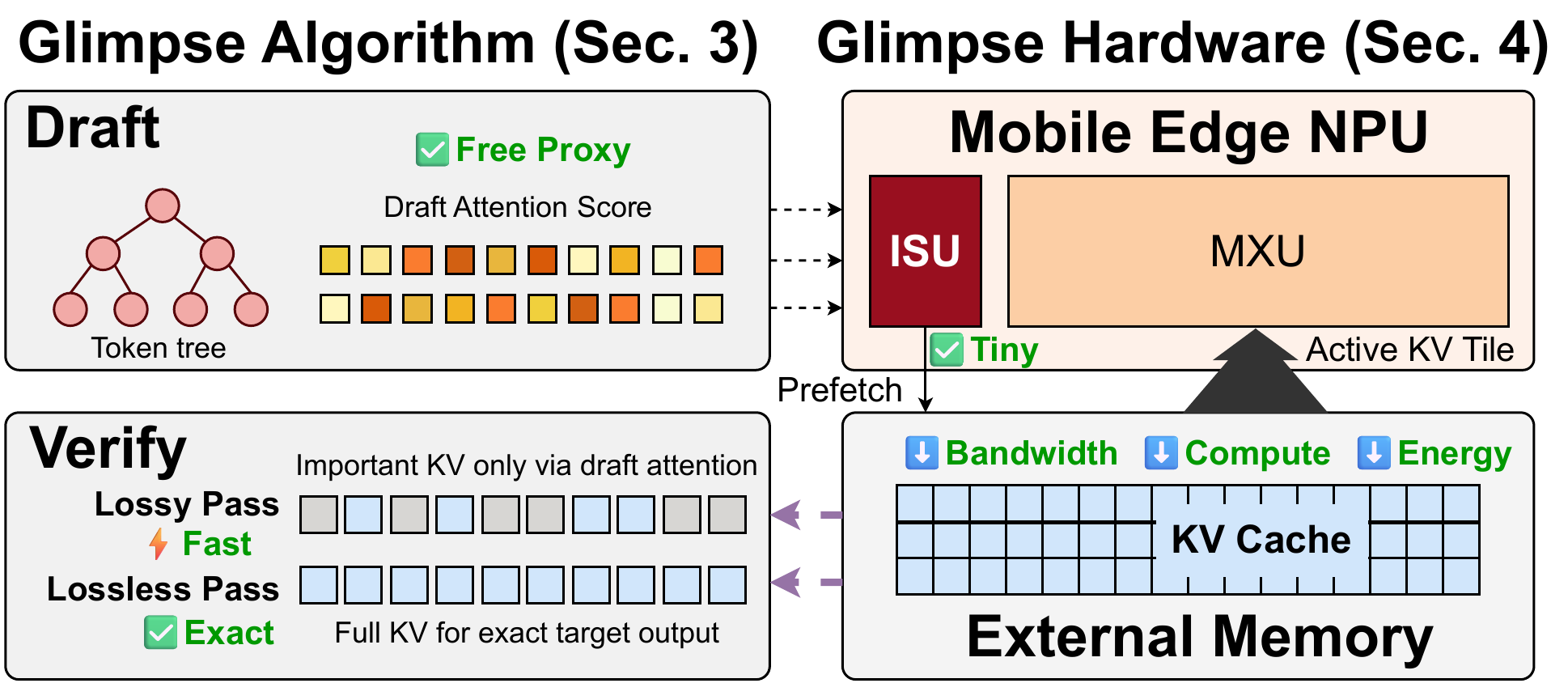}
    \vspace{-2em}
    \caption{\fcRtwo{\projName overview. The algorithm treats draft attention as a \emph{free proxy} for prefix importance, pairing a \emph{fast} lossy pass that loads only important KV with an \emph{exact} lossless pass that restores the target output. The hardware realizes this with a \emph{tiny} Importance Selection Unit (ISU) inside the edge NPU.}
    }
    \vspace{-1em}
    \label{fig:tapas_overview}
\end{figure}

\fcRone{Prior arts tackle this bottleneck from several angles, but none fully resolves the KV bandwidth problem itself.}
\fcRthree{Processing-in-memory~\cite{P3-LLM, NeuPIMs, AttAcc, AttenPIM, UM-PIM} raises bandwidth at a cost prohibitive for mass-market devices, disaggregated serving~\cite{Splitwise, DistServe, Mooncake, WindServe} leaves the KV traffic of each decode step unchanged, and KV cache compression~\cite{H2O, SnapKV, TurboQuant, SCOPE, SkipKV, StreamingLLM, SAGE-KV} cuts this traffic at the price of visible quality loss under the aggressive ratios that long contexts require~\cite{KVCompression}.}
%
%
\fcRone{Instead, speculative decoding (SD)~\cite{SD} tackles the KV cache bandwidth bottleneck without the quality-cost trade-off.}
\fcRthree{A cheap \emph{draft model} speculates several tokens ahead, and the original LLM, the \emph{target model}, verifies them all in one forward pass, such that one KV cache load serves multiple accepted tokens.}
\fcRthree{Tree-based variants~\cite{SpecInfer, Medusa, EAGLE, EAGLE-2, EAGLE-3, SpecExtend, Sequoia, Hydra} push this further, e.g., the draft model of the state-of-the-art (SotA) EAGLE-3~\cite{EAGLE-3} grows a token tree whose paths the target model verifies at once, amortizing each KV load over more tokens.}

\fcRone{However, even tree-based SD still hits a bandwidth bottleneck at the edge.}
\fcRone{Every verification step loads the full prefix KV cache, containing all context accumulated so far, from external memory to validate its candidate tokens, and this cost scales with prefix length.}
\fcRone{Our profiling (Sec.~\ref{sec:motivation}) shows that at 32K context, verification attention consumes \fcRtwo{71--77\%} of per-step latency even on \fcRtwo{the RTX 3090, a desktop-class edge GPU with 936\,GB/s bandwidth}, dominated by the bandwidth cost of full-prefix KV loading and compounded by the attention compute that grows with prefix length.}
\fcRtwo{Table~\ref{tab:hw-spec} shows DRAM bandwidth on mobile edge devices is another order of magnitude lower, making this memory-bound overhead decisive.}


\newcommand{\rot}[1]{\rotatebox[origin=l]{30}{#1}}

\begin{table}[!t]
  \centering
  \caption{\fcRtwo{Hardware capabilities from cloud to mobile edge.}}
  \vspace{-0.5em}
  \label{tab:hw-spec}
  \setlength{\tabcolsep}{3pt}
  \renewcommand{\arraystretch}{1.1}
  \resizebox{\columnwidth}{!}{%
  \begin{tabular}{lccccccccc}
    \toprule
    & \multicolumn{2}{c}{Cloud} & \multicolumn{2}{c}{\fcRtwo{Desktop Edge}} & \multicolumn{5}{c}{\fcRtwo{Mobile Edge}} \\
    \cmidrule(lr){2-3}\cmidrule(lr){4-5}\cmidrule(lr){6-10}
    & \rot{H100} & \rot{TPUv5e} & \rot{RTX4090} & \rot{RTX3090}
      & \rot{Orin Nano} & \rot{M4} & \rot{M3} & \rot{A17 Pro} & \rot{SD8Gen3} \\
    \midrule
    Use case & \multicolumn{2}{c}{Datacenter} & \multicolumn{2}{c}{Workstation}
             & Dogbot & \multicolumn{2}{c}{Laptop} & \multicolumn{2}{c}{Phone} \\
    Memory type & HBM3 & HBM2e & \multicolumn{2}{c}{GDDR6X}
                & LPDDR5 & LPDDR5X & \multicolumn{2}{c}{LPDDR5} & LPDDR5X \\
    Battery (Wh) & - & - & - & - & 236.8 & 53.8 & 52.6 & 12.7 & 19 \\
    Off-chip BW (GB/s) & 3350 & 819 & 1008 & 936 & 68 & 120 & 100 & 51.2 & 77 \\
    Throughput (TOPS)  & 1979 & 393 & 660 & 284 & 20 & 38 & 18 & 35 & 45 \\
    Power / TDP (W) & 700 & $\sim$170 & 450 & 350 & 15 & $\sim$40 & $\sim$30 & $\sim$11 & $\sim$13 \\
    \bottomrule
  \end{tabular}%
  }
  \vspace{-2em}
\end{table}



\fcRone{We observe that much of this verification bandwidth is wasted.}
\fcRone{Specifically, most prefix tokens receive negligible attention and thus contribute little to the verification outcome, yet every token costs the same to load from external memory.}
\fcRone{In this paper, we argue that the SD pipeline itself contains a natural signal to identify the important tokens.}
\fcRone{During drafting, the draft model runs full attention over the entire prefix at every expansion step, and its attention scores encode each prefix position's importance.}
\fcRone{Since drafting precedes verification, these scores are available before verification begins, offering a free proxy for importance-aware KV loading.}


\fcRone{Building on this insight, we present \projName, an algorithm-architecture co-design that leverages draft attention as a low-cost importance proxy to deliver substantial verification-stage bandwidth savings.}
\fcRone{On the algorithm side, \projName uses draft attention to drive sparse verification that loads only top-ranked prefix KV entries, reducing per-step KV bandwidth.}
\fcRtwo{The selection hence reuses a signal that drafting has already produced, unlike online schemes like SpecPV~\cite{SpecPV} rebuilding it at every verification step.}
\fcRone{Since this approximate selection may deviate from the target model's true importance ranking, \projName periodically performs a full-attention re-verification pass that restores lossless output and amortizes its bandwidth cost across tokens accepted over multiple sparse passes.}
\fcRone{On the hardware side, \projName augments a conventional edge NPU with a lightweight Importance Selection Unit (ISU) that accumulates draft attention scores and selects top-K positions on-chip.}
\fcRone{The ISU attaches to the NPU as a small auxiliary module, much smaller than the Matrix Multiply Unit (MXU), incurring negligible overall area overhead.}

The contributions of this paper are as follows:
\begin{itemize}[nosep]
    \item \fcRone{\textbf{Workload analysis and key finding (Sec.~\ref{sec:motivation}).} We perform a structured and quantified analysis of verification attention in tree-based SD, revealing that the vast majority of valid token pairs (e.g., 96.9\% in our profiling) fall within the prefix region, yet each of its tokens contributes negligibly to the attention output. This shows that a substantial part of the verification-stage bandwidth requirements can be eliminated.}

    \item \fcRone{\textbf{Two-pass verification algorithm (Sec.~\ref{sec:algorithm}).} We propose the \projName algorithm, which alternates a sparse lossy pass guided by draft attention with a full-attention lossless pass that restores output quality and amortizes its bandwidth cost.
    \projName achieves 2$\times$ higher bandwidth-normalized throughput than the SotA EAGLE-3
    across 6 models and 3 benchmarks.}

    \item \fcRone{\textbf{Lightweight hardware support (Sec.~\ref{sec:hardware}).} We design the \projName microarchitecture, augmenting a conventional edge NPU with a lightweight ISU that accumulates draft attention scores and selects top-K positions on-chip. With only 2.7\% additional area, \projName achieves on average \fcRtwo{1.33--1.74$\times$} speedup and \fcRtwo{1.40--1.76$\times$} energy efficiency over the SotA EAGLE-3 \fcRtwo{across 2K--64K contexts}.}
\end{itemize}


%% file: chapters/2-motivation.tex
\section{Background and Motivation} \label{sec:motivation}

\subsection{Tree-based Speculative Decoding} \label{sec:bg_sd}

\begin{figure}[t]
    \centering
    \includegraphics[width=\linewidth]{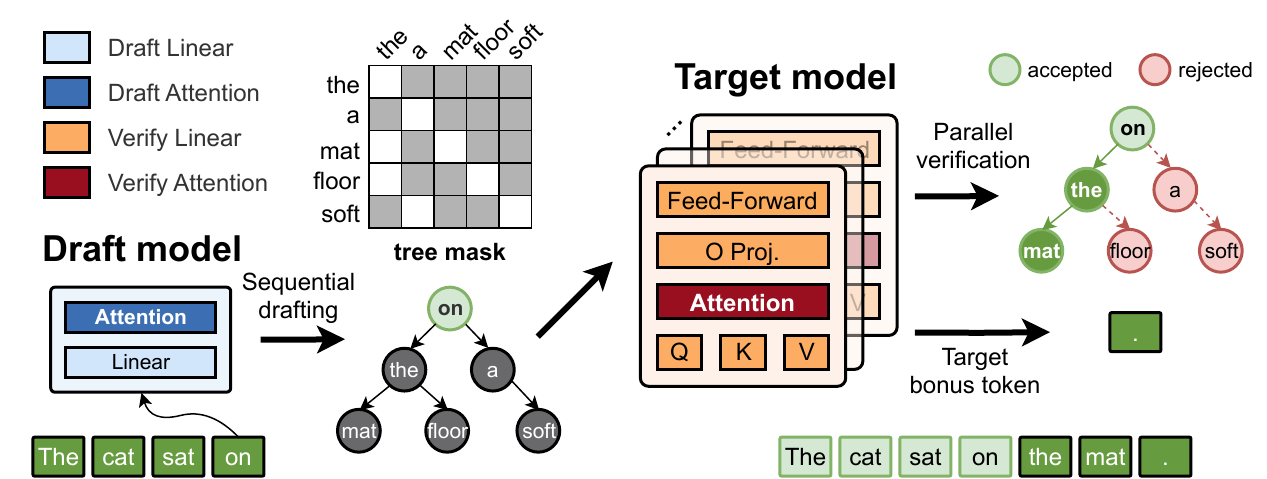}
    \vspace{-2em}
    \caption{One decode step of tree-based speculative decoding. A lightweight draft model autoregressively expands a token tree over the current prefix, and the target model then verifies all candidate paths in a single forward pass, accepting the longest matching prefix as the step output.}
    \vspace{-1em}
    \label{fig:tree_sd}
\end{figure}

\fcRone{Speculative decoding (SD)~\cite{SD, SpecAttn, SpecExtend, SpecInfer, Medusa, EAGLE, EAGLE-2, EAGLE-3, CSD, SpecExtend, Sequoia, Hydra, DSD, SpecIllusion, MagicDec, SpecEE} pairs a substantially smaller draft model (e.g., $\sim$0.5B) with a full-size target model (e.g., Llama-3.1-8B).}
\fcRone{Early SD methods~\cite{SD} verify a single chain of draft tokens per step, where one rejected token wastes all remaining candidates.}
\fcRone{Tree-based SD~\cite{SpecInfer, Medusa, EAGLE, EAGLE-2, EAGLE-3, SpecExtend, Sequoia, Hydra} addresses this by organizing candidates into a \emph{token tree}, where each root-to-leaf path represents one possible output sequence, allowing the target model to accept the longest matching path among many candidates.}
\fcRone{EAGLE-3~\cite{EAGLE-3} represents the current SotA along this line.}

\fcRone{Fig.~\ref{fig:tree_sd} illustrates the two stages of each decode step.}
\fcRone{Both draft and target models follow the standard transformer architecture~\cite{Transformer, APT-LLM}, with per-layer computation comprising linear projections and attention.}
\fcRone{In the drafting stage, the draft model autoregressively expands the token tree over the accumulated context through multiple lightweight forward passes, each producing attention scores over the full context as a natural byproduct~\cite{EAGLE, EAGLE-2, EAGLE-3}.}
\fcRone{In the verification stage, all candidate tokens are concatenated with the full context KV cache to form a unified attention input, with a \emph{tree mask} applied so that each candidate attends only to its ancestors, enforcing path independence~\cite{SpecInfer}.}
\fcRone{The target model then processes this input through all its layers, executing linear and attention operations, truncates each path at the first mismatch, and accepts the longest matching path as the step output~\cite{EAGLE-3}.}

\begin{figure}[t]
    \centering
    \includegraphics[width=\linewidth]{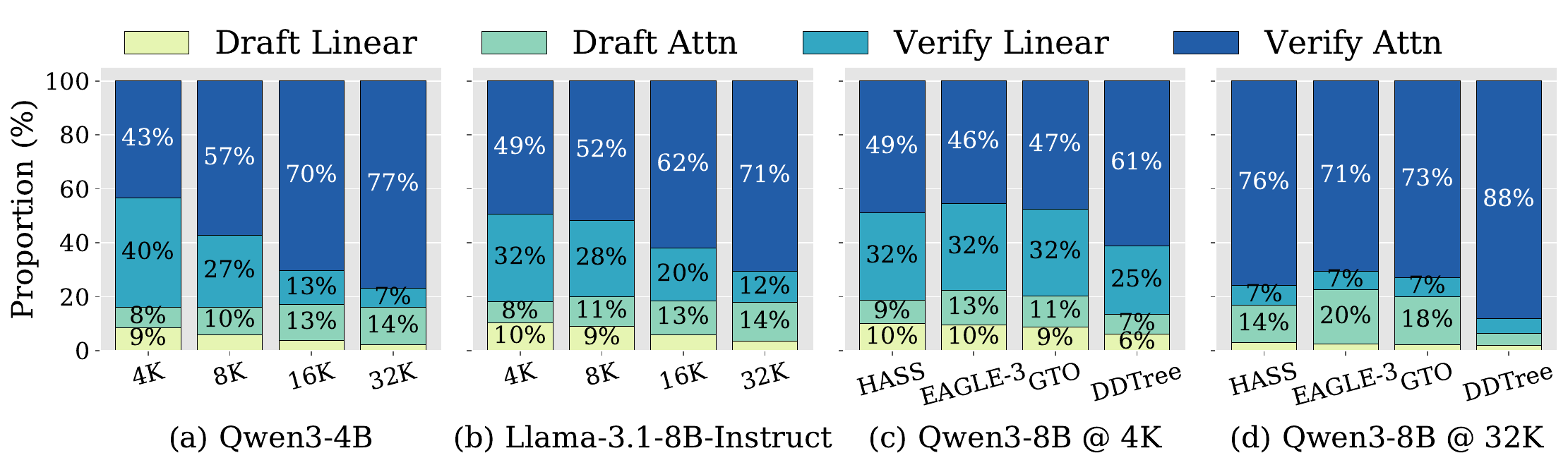}
    \vspace{-2em}
    \caption{\fcRthree{Normalized decode-step time breakdown of (a)(b) EAGLE-3 and (c)(d) four tree-based SD pipelines with different draft models and tree configurations. Verify Attention dominates increasingly with boosted contexts.}}
    \vspace{-1em}
    \label{fig:spd_profile}
\end{figure}

\subsection{Verification Bottleneck Analysis} \label{sec:profiling}

\fcRone{To understand where the cost of tree-based SD inference lies, we profile EAGLE-3~\cite{EAGLE-3} on \fcRthree{two representative models} running at batch size 1 on an NVIDIA \fcRtwo{RTX 3090 GPU~\cite{Ampere}}, using the default EAGLE-3 tree configuration\fcRthree{, and further profile four tree-based SD pipelines, HASS~\cite{HASS}, EAGLE-3, GTO~\cite{GTO}, and DDTree~\cite{DDTree}, on Qwen3-8B with their default draft models and tree configurations, as shown in Fig.~\ref{fig:spd_profile}}.}

\fcRthree{\textbf{Verify Attention becomes the bottleneck at long contexts.}}
\fcRone{We decompose each decode step into Draft Linear, Draft Attention, Verify Linear, and Verify Attention.}
\fcRthree{As shown in Fig.~\ref{fig:spd_profile}(a)(b), the lightweight draft model keeps the entire draft stage below 20\% of the step time, leaving verification as the dominant cost.
Within verification, Verify Attention takes 43--49\% at 4K, comparable to Verify Linear at 32--40\%.
At 32K it rises to 71--77\% while Verify Linear shrinks to 7--12\%, since the target model must load and attend to the entire prefix KV cache, whose memory traffic scales linearly with context, while linear-layer cost stays fixed.}

\fcRthree{\textbf{The bottleneck persists across SD pipelines.}
Fig.~\ref{fig:spd_profile}(c)(d) repeats the breakdown for four tree-based SD pipelines.
Their draft models range from the single-layer autoregressive head of EAGLE-3~\cite{EAGLE-3} to the five-layer block-diffusion drafter of DDTree~\cite{DDTree}, and their trees span 32 to 60 nodes with depths of 5 to 8.
Verification nevertheless takes at least 78\% of the step time in every pipeline, with Verify Attention reaching 71--88\% at 32K.
The bottleneck is therefore inherent to tree-based SD verification rather than specific to one draft configuration.}

\begin{figure}[t]
    \centering
    \includegraphics[width=0.95\linewidth]{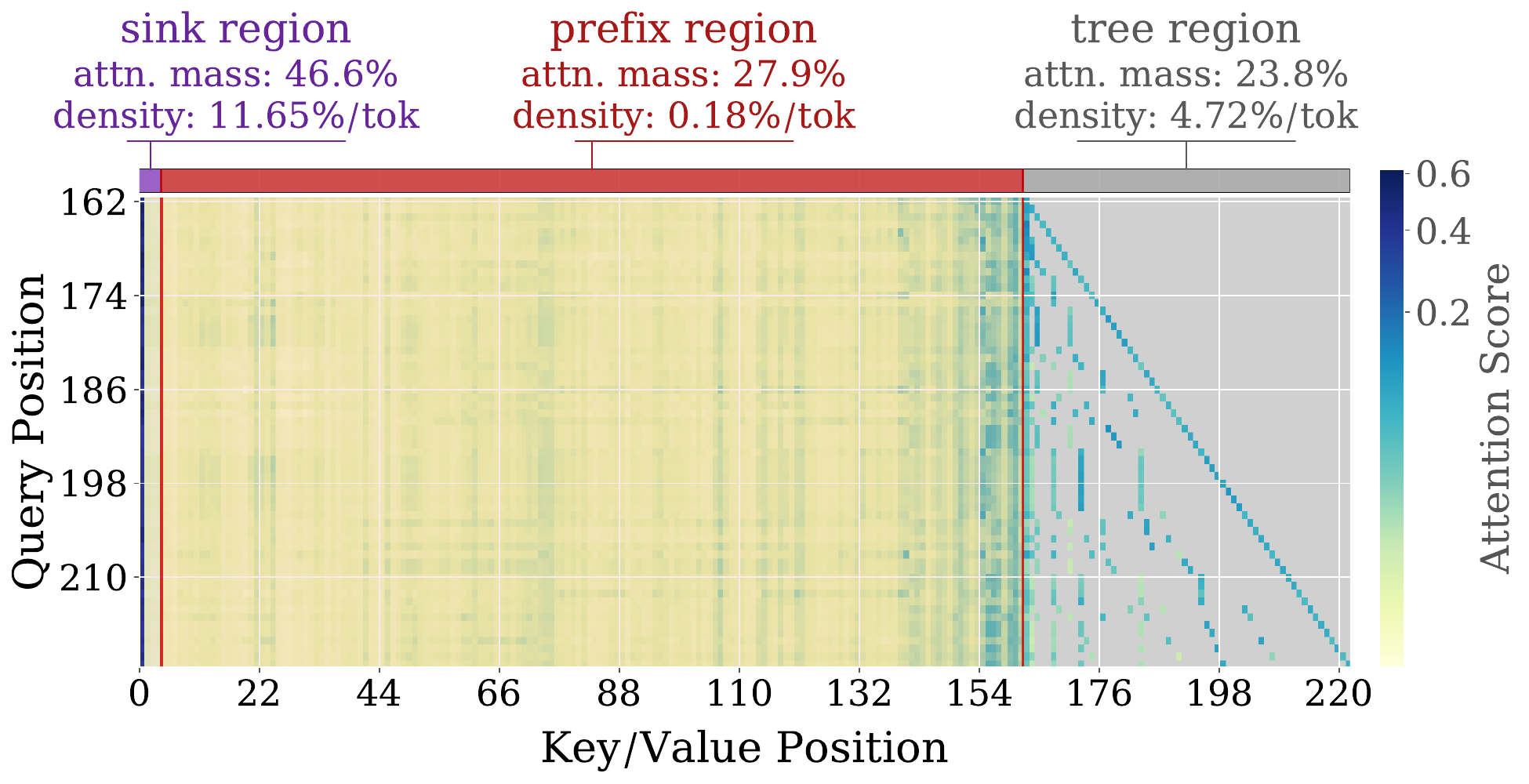}
    \vspace{-1em}
    \caption{Verification attention heatmap (Layer~15, Verify Step~15, averaged over all 32 heads) from EAGLE-3 on Llama-3.1-8B-Instruct with a GSM8K sample. The colored bar on top marks the three regions. We observe consistent patterns across different verify steps, layers, and input samples.}
    \vspace{-2em}
    \label{fig:spd_region}
\end{figure}

\textbf{Three-region decomposition.}
\fcRone{To localize this attention cost within Verify Attention, we further decompose the verification attention matrix into three regions, whose lengths we denote $L_s$, $L_p$, and $L_t$, respectively, as illustrated in Fig.~\ref{fig:spd_region}.}
\fcRone{(1)~The \emph{sink region} consists of $L_s$ \emph{sink tokens}, typically the first four, that absorb disproportionately large attention scores~\cite{StreamingLLM}.}
\fcRone{(2)~The \emph{prefix region} consists of the $L_p$ non-sink context positions, over which every tree query attends, producing $L_t \times L_p$ token pairs that constitute the dominant share of the attention computation.}
\fcRone{(3)~The \emph{tree region} consists of $L_t$ candidate tokens within the current tree, whose effective token pairs are far fewer due to the tree mask.}

\begin{figure*}[!t]
    \centering
    \includegraphics[width=\linewidth]{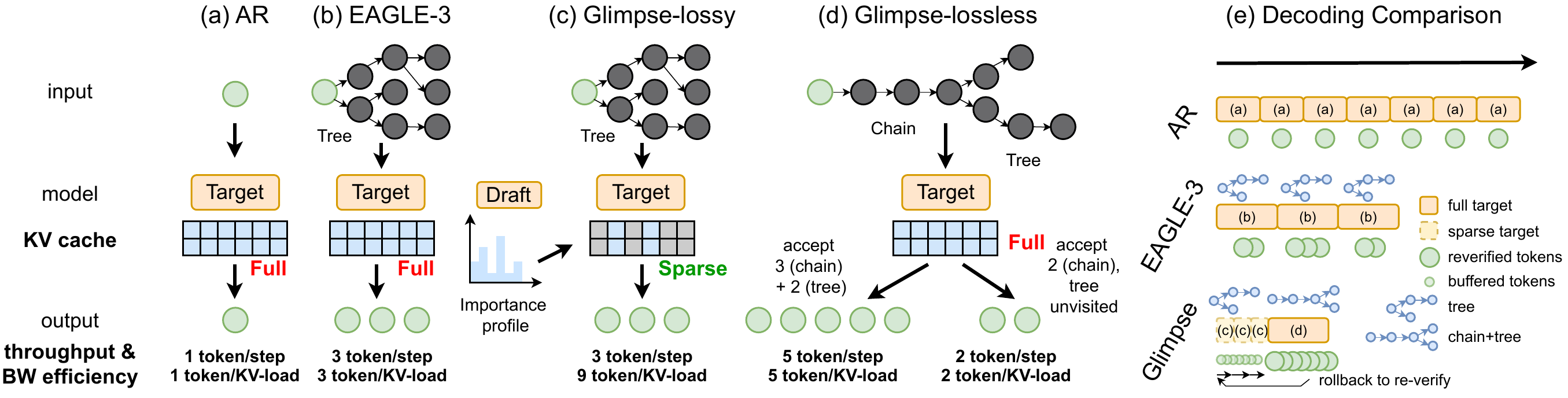}
    \vspace{-2em}
    \caption{\fcRthree{Comparison of decoding strategies from autoregressive (AR) decoding to \projName.
    (a)~\textbf{AR} verifies one token per step.
    (b)~\textbf{EAGLE-3} verifies a draft tree in one step.
    (c)~\textbf{\projName-lossy} loads only a sparse KV subset selected by the draft model's attention.
    (d)~\textbf{\projName-lossless} periodically re-verifies a merged chain + reduced-tree structure under full KV, yielding either full-chain acceptance with tree extension or early chain rejection, and always matches the target model's greedy output.
    (e)~provides a temporal view of the three decoding processes, where \projName buffers tokens across successive lossy verifications and commits them only after a periodic full-attention pass that both corrects and extends the chain.
    }}
    \vspace{-1em}
    \label{fig:algo_cmp}
\end{figure*}

\textbf{Attention mass and density.}
We define two metrics to quantify each region's importance.
\fcRone{The \emph{attention mass} of a region is the geometric mean, across all $L_t$ query tokens, of the fraction of each query's post-softmax attention allocated to that region.}
The \emph{attention density} is the attention mass divided by the number of valid key tokens in the region, representing the average attention received per token.
Intuitively, mass captures the total importance of a region, while density captures how concentrated that importance is per token.

\begin{observation}[Attention mass is highly non-uniform]
\fcRone{For this snapshot with $L_s=4$ sink tokens, $L_p=158$ prefix positions, and $L_t=60$ tree candidates, the sink region absorbs 46.6\% of the attention mass, the prefix region accounts for 27.9\%, and the tree region captures 23.8\%.}
Nearly half of attention is concentrated on a negligible number of sink tokens.
\end{observation}

\begin{observation}[Attention density reveals a stark efficiency gap]
\fcRone{The sink and tree regions reach densities of 11.65\%/tok and 4.72\%/tok, 65$\times$ and 26$\times$ denser than the prefix region's 0.18\%/tok.}
Yet the prefix region accounts for 96.9\% of all valid token pairs, dominating bandwidth consumption.
\fcRone{\emph{In other words, the vast majority of memory bandwidth is spent loading prefix KV entries whose per-token contribution to the attention output is negligible.}}
\end{observation}



\subsection{Motivation for \projName} \label{sec:design_opp}
\fcRtwo{Table~\ref{tab:hw-spec} further shows that edge platforms span from desktop GPUs to mobile SoCs. The RTX 3090 used in the above analysis enjoys 936\,GB/s of GDDR6X bandwidth~\cite{Ampere}, whereas mobile edge devices~\cite{EdgeLLM, LPDDR5} offer only 51--120\,GB/s, another order of magnitude lower. Their on-chip SRAM, only a few MB, cannot hold even a single layer's KV cache. Streaming it from off-chip memory at every verification step therefore becomes the decisive obstacle for mobile edge SD deployment.}

\fcRone{Addressing this bottleneck on the edge motivates \projName's design along three properties. First, verification should be \emph{importance-aware}, loading KV entries in proportion to their contribution rather than uniformly. Second, the importance signal should be \emph{available before verification} to steer KV loading from the start. Third, the hardware realization should be \emph{lightweight} to fit within edge NPU area and power budgets.}
%
\fcRthree{The SD pipeline itself satisfies all three properties, since the draft model already attends to the whole context before verification, and its scores form a one-dimensional profile amenable to lightweight on-chip processing.}
\fcRone{We present the resulting two-pass algorithm in Sec.~\ref{sec:algorithm} and the on-chip unit realizing it in Sec.~\ref{sec:hardware}.}








%% file: chapters/3-algorithm.tex
\section{\projName Algorithm} \label{sec:algorithm}

\subsection{Draft Attention as Importance Proxy} \label{sec:attn_proxy}

\fcRone{\fcRthree{We now quantitatively verify the draft-attention signal identified in Sec.~\ref{sec:design_opp}.}
Although a single-layer draft network is much lighter than the multi-layer target, accumulating its attention scores across multiple drafting steps yields a stable profile that highlights consistently attended prefix positions, which we expect to align with the target's importance ranking.}

\begin{figure}[t]
    \centering
    \includegraphics[width=\linewidth]{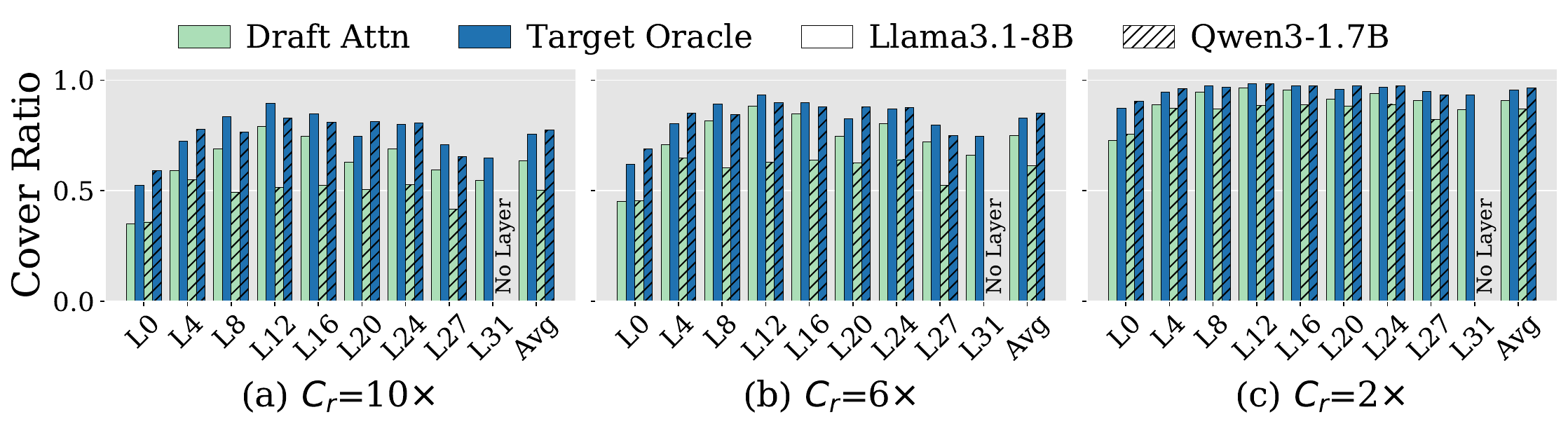}
    \vspace{-2em}
    \caption{\fcRthree{Per-layer attention coverage of Draft Attention and Target Oracle on Llama3.1-8B and Qwen3-1.7B at three compression ratios, averaged over all verification steps of 50 GSM8K~\cite{GSM8K} samples.}}
    \vspace{-1em}
    \label{fig:draft_proxy}
\end{figure}

\begin{figure}[t]
    \centering
    \includegraphics[width=\linewidth]{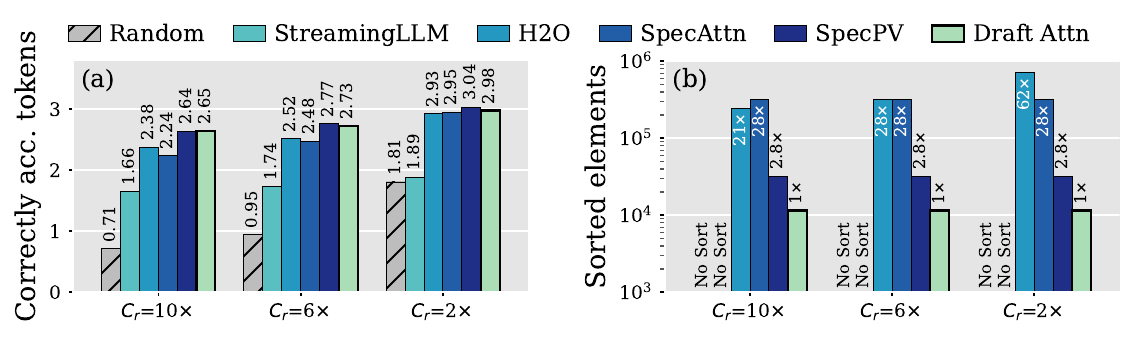}
    \vspace{-2em}
    \caption{\fcRthree{Correctly accepted tokens per step (a) and sorted elements per step relative to Draft Attn (b) of six KV selection policies under EAGLE-3 sparse verification on Qwen3-1.7B with 30 LongBench~\cite{LongBench} documents.}}
    \vspace{-1em}
    \label{fig:proxy_cmp}
\end{figure}

\fcRone{\fcRthree{We run EAGLE-3 on two target models from different families, Llama-3.1-8B~\cite{Llama3} and Qwen3-1.7B~\cite{Qwen3}, whose single-layer drafts have about 5\% and 8\% of the target parameters, respectively.}
For each verification step, we accumulate the draft model's attention scores across all drafting steps into a one-dimensional prefix importance profile, and record the target model's per-layer attention distribution averaged across heads and tree query tokens.
We define the \emph{Attention Coverage Ratio}.
When loading only the Top-$K$ prefix positions ranked by the draft profile, the coverage at target layer~$l$ is}
\[ \mathrm{Coverage}(K, l) = \sum\nolimits_{i \in \mathrm{TopK_{draft}}} a_i^{(l)} \big/ \sum\nolimits_{j=1}^{L_p} a_j^{(l)}, \]
\fcRone{where $a_i^{(l)}$ is the average target attention weight allocated to prefix position~$i$ in layer~$l$, and the denominator sums over the $L_p$ prefix positions such that Coverage measures how much of the layer's prefix-region attention mass is captured by the draft's Top-$K$.
\fcRthree{As a ground-truth upper bound that is unattainable in practice, we compare against a \emph{Target Oracle} that selects Top-$K$ from each layer's own full-attention ranking.}}
\fcRthree{Fig.~\ref{fig:draft_proxy} reports per-layer coverage at $C_r \in \{10\times, 6\times, 2\times\}$.
Even at $C_r{=}10\times$, the Target Oracle captures most of the attention mass, indicating that a small subset of prefix positions suffices to preserve importance.
On both models, Draft Attention follows the layer-wise pattern of the oracle, and its gap narrows as $C_r$ relaxes, reaching 87--91\% average coverage at $C_r{=}2\times$.}

\fcRthree{We further examine how many draft tokens each step accepts correctly, which directly determines the verification yield.
Since sparse verification may accept tokens that full attention would reject, a token counts as correctly accepted only if it matches the full-attention verification of the same draft tree.
To isolate the importance signal, Fig.~\ref{fig:proxy_cmp} applies Draft Attention and the random, StreamingLLM~\cite{StreamingLLM}, H2O~\cite{H2O}, SpecAttn~\cite{SpecAttn}, and SpecPV~\cite{SpecPV} selection policies to the same EAGLE-3 sparse verification under the same KV budget.
As Fig.~\ref{fig:proxy_cmp}(a) shows, Draft Attention matches SpecPV, as both score the full prefix with the tokens about to be verified.
At $C_r{=}10\times$, Draft Attention accepts 11--60\% more correct tokens than the other three, which rely on recency or past attention and miss tokens that become important later.
Fig.~\ref{fig:proxy_cmp}(b) counts the elements entering Top-K sorts per step.
SpecPV sorts 2.8$\times$ as many elements as Draft Attention, and the other ranking policies 21--62$\times$ as many, since they rank the prefix per layer or per KV head with scores from verification, whereas Draft Attention sorts a single profile produced during drafting.
Draft Attention is thus an importance proxy as accurate as SpecPV, yet with the fewest sorted elements among ranking policies.}


\begin{algorithm}[t]
\footnotesize 
\caption{\projName Two-Pass Speculative Decoding}\label{alg:tapas}
\DontPrintSemicolon
\SetAlgoLined
\KwIn{prompt $x$, target model $M_T$, draft model $M_D$, compression ratio $C_r$, chunk size $C_s$, verification depth $T_y$}
\KwOut{generated token sequence $y$}
$\mathit{ckpt} \leftarrow |x|$\;
\While{$\mathrm{EOS} \notin y$ \textnormal{and} $|y| < |y|_{\max}$}{
    $\mathit{buf} \leftarrow [\,]$\;
    \tcc{\textbf{Lossy Phase}}
    \While{$|\mathit{buf}| < T_y$}{
        $\mathit{tree},\; \mathit{attn\_profile} \leftarrow \textsc{EagleDraft}(M_D)$\;
        $\mathit{chunks} \leftarrow \textsc{ChunkPool}(\mathit{attn\_profile},\; C_s)$\;
        $K \leftarrow \lfloor L_p / (C_r \cdot C_s) \rfloor$\;
        $\mathit{top\text{-}K} \leftarrow \textsc{TopK}(\mathit{chunks},\; K)$\;
        $\mathit{mask} \leftarrow \mathit{sink} \cup \textsc{Expand}(\mathit{top\text{-}K}) \cup \mathit{tree}$\;
        $\mathit{accepted} \leftarrow \textsc{TreeVerify}(M_T,\; \mathit{tree},\; \mathit{mask})$ \fcRthree{\textit{// sparse}}\;
        $\mathit{buf}.\mathrm{append}(\mathit{accepted})$\;
    }
    \tcc{\textbf{Lossless Phase}}
    Rollback KV cache to $\mathit{ckpt}$\;
    $\mathit{tree}_r \leftarrow \textsc{EagleDraft}(M_D,\; \mathrm{reduced})$\;
    $\mathit{merged} \leftarrow \textsc{MergeChainTree}(\mathit{buf},\; \mathit{tree}_r)$\;
    $\mathit{accepted} \leftarrow \textsc{TreeVerify}(M_T,\; \mathit{merged},\; \mathrm{full})$ \fcRthree{\textit{// full}}\;
    $y \leftarrow y \cup \mathit{accepted}$\;
    $\mathit{ckpt} \leftarrow$ new context length\;
}
\Return $y$\;
\end{algorithm}

\subsection{Chunk-Pooled Lossy Verification} \label{sec:lossy_pass}

\fcRone{\fcRthree{On top of the EAGLE-3 pipeline, the lossy pass turns the draft model's attention into a prefix \emph{importance profile} used for selective KV loading, as illustrated in Fig.~\ref{fig:algo_cmp}(c) and Algorithm~\ref{alg:tapas}.}
At each drafting step, \projName accumulates the draft's attention scores over the prefix positions and normalizes them across drafting steps within a single draft round, yielding a one-dimensional profile that estimates each prefix position's contribution to the verification.}

\fcRone{Selecting individual prefix positions from this profile would issue scattered, non-contiguous KV loads, wasting most of each memory burst and eroding the intended bandwidth saving.
\projName therefore partitions the prefix into fixed-size chunks of $C_s$ contiguous tokens, scores each chunk by the mean of its token importances, and performs Top-$K$ selection at chunk granularity.
The number of chunks loaded is governed by a compression ratio $C_r$ as $K = \lfloor L_p / (C_r \cdot C_s) \rfloor$.
}

\fcRone{During lossy verification, \projName loads three KV regions.
The $L_s$ sink tokens and the $L_t$ tree candidates are always loaded, since Observation~2 and Fig.~\ref{fig:spd_region} have shown that both regions carry high attention density despite their small token counts.
For the $L_p$ prefix region, which has the lowest density but occupies most of the memory footprint, \projName loads only the chunks selected by the draft profile and leaves the rest skipped.
}

\begin{figure}[t]
    \centering
    \vspace{-1em}
    \includegraphics[width=0.95\linewidth]{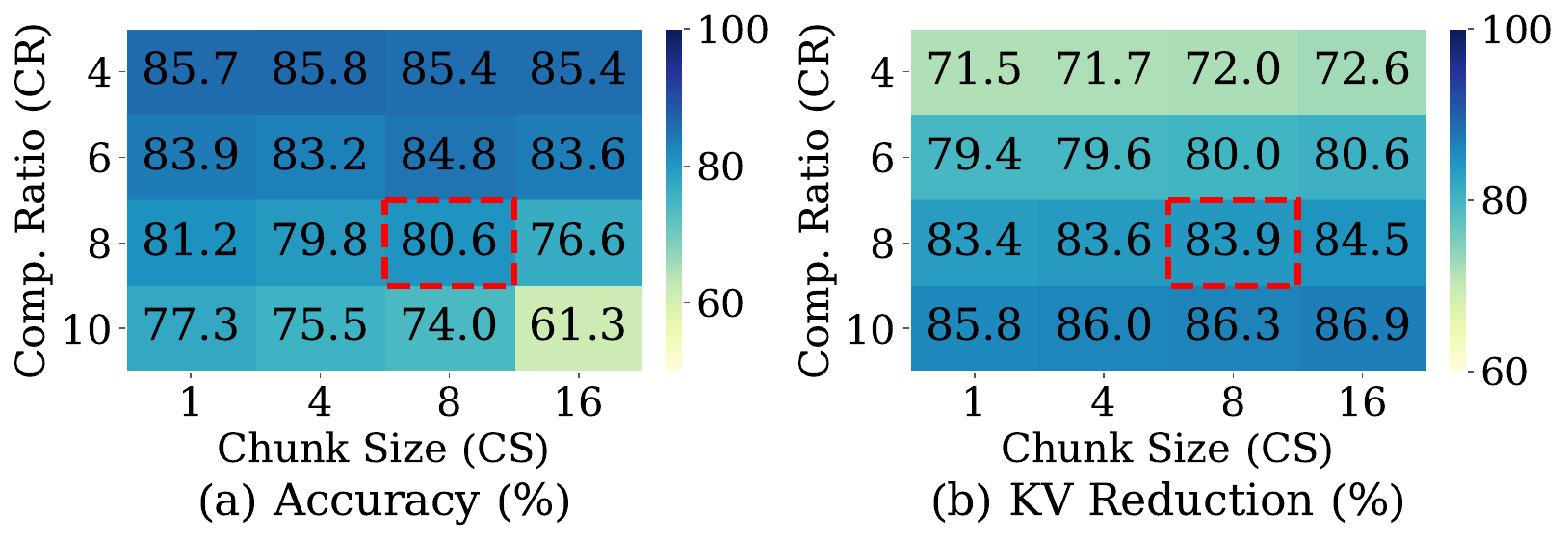}
    \vspace{-1em}
    \caption{\fcRthree{(a) Accuracy and (b) KV reduction under varying $C_r$ and $C_s$ on Llama3.1-8B with GSM8K lossy-only mode. The dashed box marks the selected operating point ($C_r{=}8$, $C_s{=}8$).}}
    \vspace{-1em}
    \label{fig:cr_cs_tradeoff}
\end{figure}

\fcRone{The lossy pass introduces two hyperparameters, $C_r$ and $C_s$.
Fig.~\ref{fig:cr_cs_tradeoff} sweeps both on \fcRthree{Llama3.1-8B} with GSM8K under lossy-only mode to characterize their joint effect.
$C_r$ dominates the accuracy-compression tradeoff, while $C_s$ has marginal impact, with accuracy drop $<2\%$ across $C_s \in \{1,4,8,16\}$.
We adopt $C_r{=}8$ and $C_s{=}8$ as defaults.
Although $C_r{=}8$ alone incurs a visible accuracy drop, the lossless re-verification pass recovers it, allowing the aggressive compression that maximizes bandwidth savings.}


\subsection{Full-Attention Lossless Re-Verification} \label{sec:lossless_pass}

\fcRone{The \emph{lossless} pass periodically inserts a re-verification with full attention to recover exact output quality.
\fcRthree{At aggressive $C_r$, however, the sparse attention of the lossy pass can alter accept/reject decisions and drift from the target output, as Fig.~\ref{fig:cr_cs_tradeoff} shows.}}
\fcRone{Each sparse step appends accepted tokens to a buffer.
The checkpoint $\mathit{ckpt}$ tracks the KV cache length corresponding to the committed output $y$.
When the buffer length reaches the \emph{verification speculation depth} $T_y$, \projName rolls back the KV cache to this checkpoint, discarding all KV entries written under incomplete attention.
A sequential chain verification would re-check the lossy-accepted chain token by token under full attention, but this forfeits tree-based SD's branching and caps the step's yield at the chain's contiguous correct prefix.
\projName instead merges the chain with a freshly drafted \emph{reduced tree} of size $T_r = \max(L_t+1-|\mathit{buf}|,\,1)$ into a unified structure, keeping the merged length matched to a standard verification step such that the lossless pass costs no more than a regular one.
As illustrated in Fig.~\ref{fig:algo_cmp}(d) \fcRthree{and Algorithm~\ref{alg:tapas}}, this merged structure allows the lossless pass to either extend the accepted chain with further tree candidates or detect a rejection early in the chain.}


\fcRthree{Under greedy decoding, the lossless pass verifies the merged structure under full attention over the complete prefix KV cache, accepts the longest prefix whose tokens match the target model's argmax predictions, and takes the target model's own prediction at the first mismatch.}
\fcRtwo{The lossy pass therefore determines which tokens are verified, while the lossless pass determines which are accepted, and only the latter affects the output.}
\fcRthree{Every committed token is thus identical to the target model's greedy autoregressive output, regardless of the KV that the lossy pass selects.
Under stochastic sampling, the lossless pass would follow the tree verification of EAGLE-3~\cite{EAGLE-3}, which accepts each candidate $x$ with its full-attention target probability $p(x)$ and otherwise removes $x$ from $p$ and resamples.
This rule does not involve draft probabilities and thus preserves the target distribution for any proposal, and the buffered tokens need no proposal probability.}

\begin{figure}[t]
    \centering
    \includegraphics[width=\linewidth]{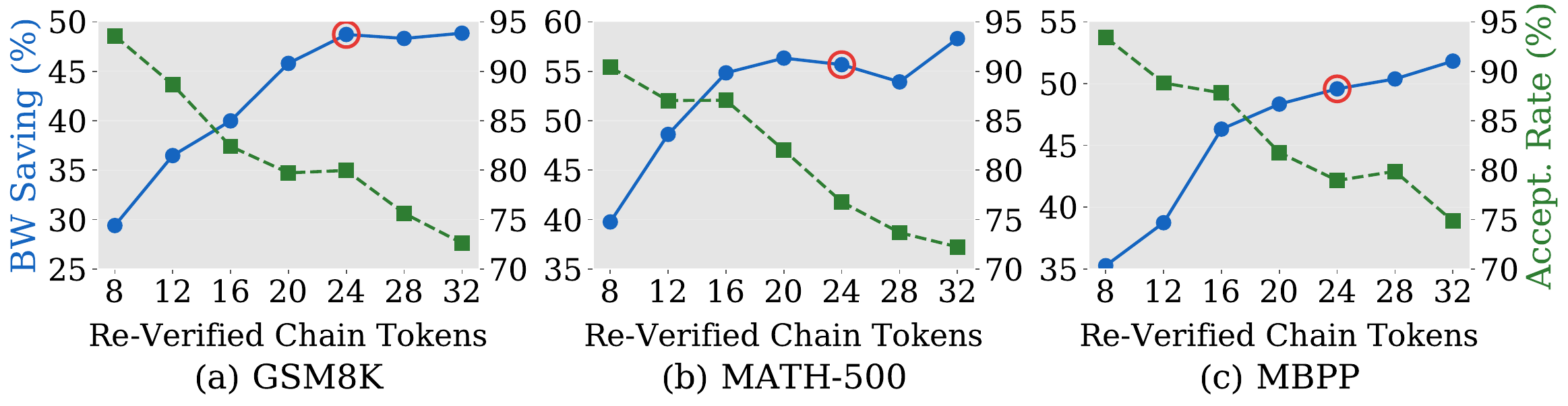}
    \vspace{-2em}
    \caption{\fcRthree{Bandwidth saving over EAGLE-3 full-prefix-load verification derived from $\tau_\mathrm{bw}$ (left) and lossless acceptance rate (right) under varying $T_y$ on Llama3.1-8B. The circled point marks the selected default $T_y{=}24$.}}
    \label{fig:ty_sensitivity}
    \vspace{-1em}
\end{figure}

\fcRone{$T_y$ governs the balance between the two phases and is the key efficiency knob.
Standard SD measures efficiency by the mean accepted tokens per verification step,}
\[ \tau_\mathrm{step} = N_\mathrm{tokens} / N_\mathrm{steps}, \]
\fcRone{which treats every step as having equal cost.
In \projName, however, a lossy step loads only compressed KV while a lossless step loads the full cache, an asymmetry $\tau_\mathrm{step}$ cannot capture.
Let $N_\mathrm{lossy}$ and $N_\mathrm{lossless}$ denote the total counts of lossy and lossless verification steps during generation, and let $\rho_\mathrm{kv} \in [0,1]$ be the measured fraction of prefix KV skipped in a lossy step.
We then define a \emph{bandwidth-normalized token throughput}}
\[ \tau_\mathrm{bw} = \frac{N_\mathrm{tokens}}{N_\mathrm{lossy}(1-\rho_\mathrm{kv}) + N_\mathrm{lossless}}, \]
\fcRone{which weights each step by its actual bandwidth cost.
Note that $\rho_\mathrm{kv}$ is close to but not exactly $1-1/C_r$ due to sink and tree-region overhead.
Bandwidth saving, reported in Fig.~\ref{fig:ty_sensitivity}, is measured relative to EAGLE-3's full-prefix-load verification.}

\fcRone{To choose $T_y$, Fig.~\ref{fig:ty_sensitivity} sweeps it on \fcRthree{Llama3.1-8B} ($C_r{=}8$, $C_s{=}8$) for three tasks.
\fcRthree{Larger $T_y$ raises the share of lossy steps and thus the bandwidth saving.}
Meanwhile the lossless acceptance rate decays monotonically.
As the lossy chain grows longer, approximation errors from sparse attention compound across successive accept/reject decisions, and each additional step is more likely to diverge from full attention, increasing the chance of an early rejection during re-verification.
This trend corroborates the need for periodic lossless correction, since a purely lossy chain would keep degrading without rollback.}

\fcRthree{\subsection{Trace Analysis and Comparison with Prior SD Methods} \label{sec:trace_analysis}}

\begin{figure}[t]
    \centering
    \includegraphics[width=\linewidth]{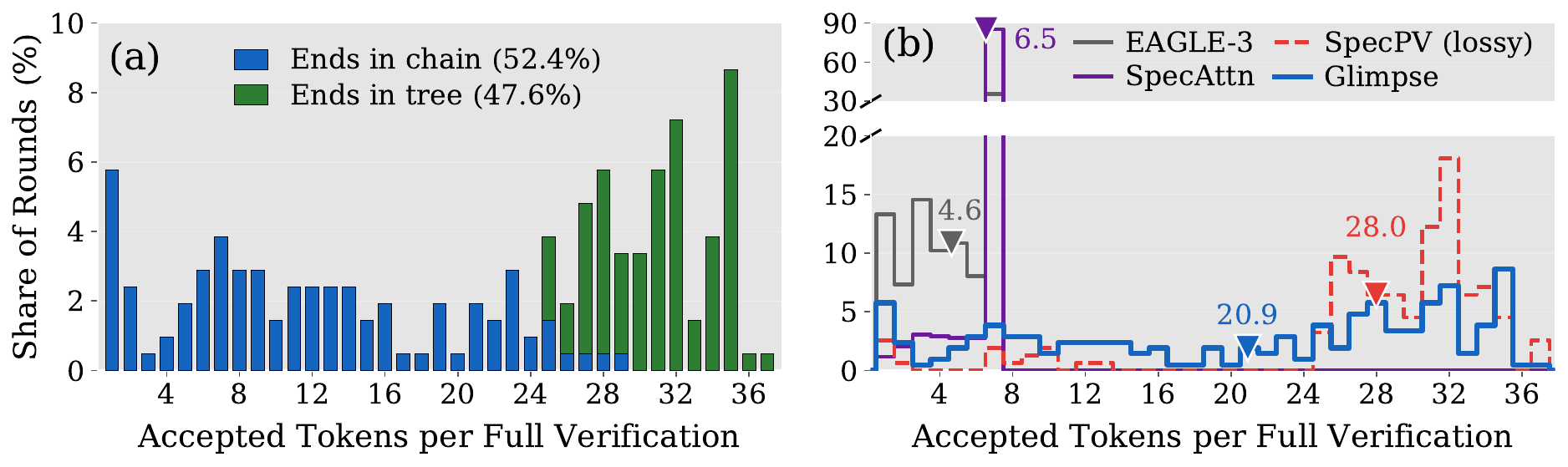}
    \vspace{-2em}
    \caption{\fcRthree{Accepted tokens per full verification on Llama3.1-8B~\cite{Llama3} with 10 samples in MATH500~\cite{MATH500}. (a)~\projName rounds that end in the lossy chain or extend into the merged tree. (b)~Comparison with EAGLE-3~\cite{EAGLE-3}, SpecAttn~\cite{SpecAttn}, and SpecPV~\cite{SpecPV}. Triangles mark the mean accepted tokens.}}
    \vspace{-1em}
    \label{fig:trace_analysis}
\end{figure}

\fcRthree{Buffering $T_y$ lossy tokens pays off only if most of the chain survives the full-attention check.
To illustrate how the lossless pass behaves, Fig.~\ref{fig:trace_analysis} traces the tokens accepted per full verification, i.e., per load of the entire prefix KV cache.}

\fcRthree{Fig.~\ref{fig:trace_analysis}(a) shows where each full verification stops.
Only 8.2\% of the rounds stop within the first two tokens, where most of the buffered chain is discarded, while the other rejections are spread along the chain and keep all the tokens before them.
At the other end, 47.6\% of the rounds extend beyond the chain into the merged tree.
When the chain survives, nothing is discarded and the tree adds 4.6 tokens on average, a gain that chain-only correction would forgo.
The rollback penalty, which consists of sparse-KV steps only, is therefore outweighed by the tokens that the surviving chains and their tree extensions commit.}

\fcRthree{Fig.~\ref{fig:trace_analysis}(b) compares \projName to prior SD methods.
EAGLE-3~\cite{EAGLE-3} performs a full KV load at every step, and its accepted tokens are bounded by the tree depth.
SpecAttn~\cite{SpecAttn} is a self-speculative scheme that drafts with the target model itself under sparse attention.
Its heavy draft is accurate and most chains are accepted in full, which produces the peak in Fig.~\ref{fig:trace_analysis}(b), yet each chain is short and the full KV cache is reloaded every few tokens.
SpecPV~\cite{SpecPV}, also self-speculative, verifies with a partial KV cache and amortizes one full KV load over many tokens, but it commits the buffered chain without re-verification, and its output deviates from that of the target model.
Prior methods thus either stay exact by performing a full KV load every few tokens, or amortize it at the expense of exactness.
In contrast, \projName separates the two roles.
Sparse verification only proposes tokens, and the amortized full-attention pass alone commits them.
On average, \projName commits 20.9 tokens per full KV load, against 4.6 for EAGLE-3 and 6.5 for SpecAttn, while SpecPV reaches 28.0 only without re-verification.
\projName therefore trades the KV traffic of its lossy steps against the number of tokens that each full KV load commits, and leaves the output identical to the target model's greedy autoregressive output.}

%% file: chapters/4-hardware.tex
\vspace{8pt}\kern0pt 
\section{\projName Architecture} \label{sec:hardware}

\subsection{Overall Architecture} \label{sec:hw_overview}

\fcRone{Fig.~\ref{fig:tapas_arch}(a) shows the baseline edge NPU~\cite{SamsungNPU, NPU-LLM, tpuv4i, tpu, Mokey, ZigZag, ZigZagLLM} that \projName builds upon.
It comprises a $64\times64$ output-stationary matrix multiply unit (MXU), a special function unit (SFU) for nonlinear operations such as softmax, three on-chip SRAMs of 2\,MB Weight, 2\,MB Activation, and 128\,KB Output each organized as ping-pong dual banks, together with a Memory Controller managing off-chip DRAM traffic.
Modules are connected through the SRAM buffers, with the MXU executing all matrix multiplications and the SFU handling softmax and other element-wise nonlinearities.}
\fcRone{The MXU supports a W4A16 mixed-precision mode (4-bit weights with 16-bit activations)~\cite{GPTQ, AWQ, OmniQuant, Anda, FIGNA, BitMoD, COMET}, in which 4-bit integer weights are decoded to BF16 by a lightweight INT4-to-BF16 conversion unit at the processing element (PE) input and fed into the BF16 multiply-accumulate (MAC) datapath.
This matches the de facto precision convention for edge LLM deployment~\cite{apple, mlc-llm}, where linear layers adopt 4-bit weight quantization while the KV cache is kept in native BF16.
Both draft and verify stages use this split.}
\fcRone{\projName attaches the ISU to the baseline NPU through a tap on the Output SRAM's write-out path and a control channel to the Memory Controller.
The baseline datapath is left unchanged, and the NPU behaves identically to the original design whenever the ISU is inactive.
}

\subsection{Importance Selection Unit} \label{sec:hw_isu}

\begin{figure}[t]
    \centering
    \includegraphics[width=\linewidth]{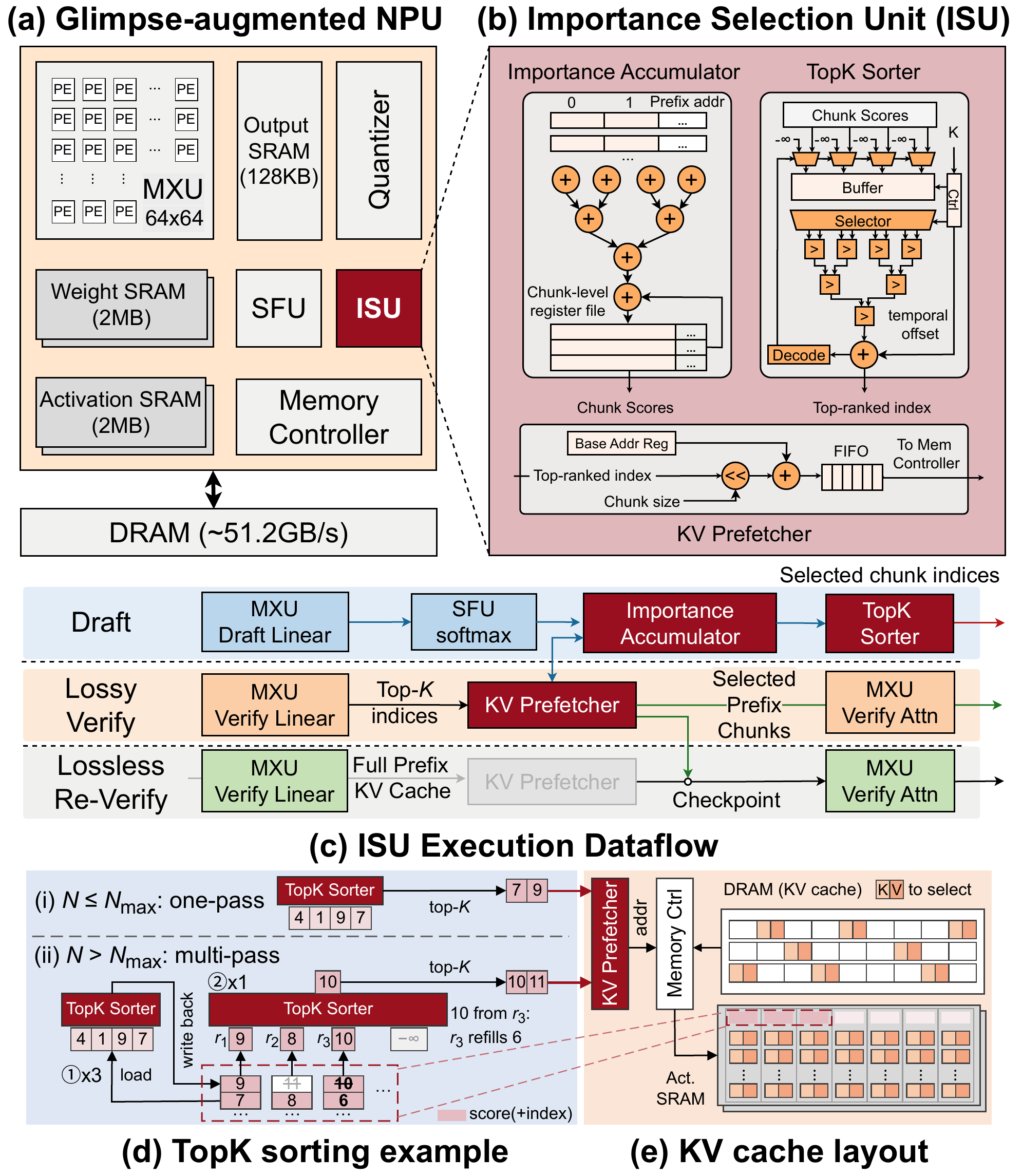}
    \vspace{-2em}
    \caption{\fcRthree{\projName hardware architecture. (a)~Overall NPU block diagram with the added ISU highlighted. (b)~ISU internal structure with its Importance Accumulator, TopK Sorter, and KV Prefetcher. (c)~ISU execution dataflow across the three phases. (d)~TopK sorting example. (e)~KV cache layout.}}
    \label{fig:tapas_arch}
    \vspace{-1em}
\end{figure}

\fcRone{Fig.~\ref{fig:tapas_arch}(b) shows the internal structure of the ISU, which consists of three modules.
The Importance Accumulator consumes softmax outputs from the SFU and builds a chunk-level importance profile.
The TopK Sorter selects the $K$ most important chunks from this profile.
The KV Prefetcher translates the selected chunk indices into DRAM addresses and drives the Memory Controller to issue KV cache loads.}

\fcRone{The Accumulator exploits an observation made in Sec.~\ref{sec:lossy_pass}, that the two reduction axes of the attention profile, temporal reduction across expansion steps and spatial reduction across the $C_s$ positions within one chunk, can share the same hardware resources once the register file itself is organized at chunk granularity.
Each Accumulator entry corresponds to one chunk of $C_s$ contiguous prefix positions rather than an individual token.
When a post-softmax attention weight arrives, its prefix position is mapped to a chunk entry via a simple address translation, and a read-modify-write accumulation is performed on that entry.
Scores from adjacent positions within one expansion step land on the same entry, realizing spatial reduction, and scores from different expansion steps add to the same set of entries over time, realizing temporal reduction.
Both reductions share one adder and one register file, and the Accumulator's storage scales with the number of chunks rather than the prefix length.}

\fcRone{The chunk-level profile produced by the Accumulator is loaded into the leaf nodes of the TopK Sorter, which employs a tournament tree~\cite{TAOCP} supporting up to $N_{\max} = 2048$ chunk candidates for one time.
Each internal node stores not only the maximum score of its subtree but also the identity of the winning leaf, such that the root directly reveals both the top chunk and its index without walking the tree.
Each extraction round invalidates the winning leaf by setting its score to $-\infty$, and the Sorter restores correctness by re-comparing only along the single path from that leaf to the root, avoiding a full rebuild.
$K$ rounds of extraction therefore take time proportional to $K \log N$, and the Sorter consists only of comparators and multiplexers.}
\fcRthree{When the prefix exceeds $N_{\max}$ chunks, the Sorter runs in two phases, as Fig.~\ref{fig:tapas_arch}(d) illustrates with $N_{\max}{=}4$, $K{=}2$, and 12 chunks.
It first sorts each run of $N_{\max}$ chunks and writes its top-$K$ back to the Activation SRAM, laid out as in Fig.~\ref{fig:tapas_arch}(e), and then merges the runs by loading their heads into the leaves and refilling each extracted leaf from its run.
Each extraction or refill updates a single leaf-to-root path, and the two phases thus take $\lceil N/N_{\max}\rceil + 1$ times the $K \log N_{\max}$ cycles of a single pass.
Even this multi-pass cost remains negligible next to the KV loading that the selection avoids.}

\fcRone{
\projName lays out each chunk as a contiguous address range in DRAM and restricts selection to chunk granularity, such that each chunk's DRAM start address follows directly from its index through a single address-generation step, requiring no scatter-gather logic.
The $K$ resulting addresses are queued into the Memory Controller's DMA request queue in order, and the Prefetcher itself contains only one address-generation datapath and a small request FIFO.}
\fcRthree{The original DMA thus remains the only bus master and serves these requests after weight requests, using only idle bus cycles.}

\subsection{Execution Flow} \label{sec:hw_dataflow}

\begin{table*}[t]
\centering
\caption{\fcRthree{Algorithm-level comparison across six models and three benchmarks. All three metrics are higher-is-better: accuracy (\%), $\tau_\mathrm{step}$ measures accepted tokens per decoding step, and $\tau_\mathrm{bw}$ measures tokens generated per full KV cache access.}}
\label{tab:algo_main}
\resizebox{\textwidth}{!}{%
\begin{tabular}{llccc|ccc|ccc|ccc|ccc|ccc}
\toprule
\multicolumn{2}{c}{} & \multicolumn{3}{c}{\textbf{DeepSeek-R1-8B}} & \multicolumn{3}{c}{\textbf{Llama3.1-8B}} & \multicolumn{3}{c}{\textbf{Qwen3-1.7B}} & \multicolumn{3}{c}{\textbf{Qwen3-4B}} & \multicolumn{3}{c}{\textbf{Qwen3-8B}} & \multicolumn{3}{c}{\textbf{Qwen3-14B}} \\
\cmidrule(lr){3-5} \cmidrule(lr){6-8} \cmidrule(lr){9-11} \cmidrule(lr){12-14} \cmidrule(lr){15-17} \cmidrule(lr){18-20}
\textbf{Dataset} & \textbf{Method} & Acc & $\tau_\mathrm{step}$ & $\tau_{bw}$ & Acc & $\tau_\mathrm{step}$ & $\tau_{bw}$ & Acc & $\tau_\mathrm{step}$ & $\tau_{bw}$ & Acc & $\tau_\mathrm{step}$ & $\tau_{bw}$ & Acc & $\tau_\mathrm{step}$ & $\tau_{bw}$ & Acc & $\tau_\mathrm{step}$ & $\tau_{bw}$ \\
\midrule
\multirow{6}{*}{GSM8K} & AR & 87.8 & 1.00 & 1.00 & 85.7 & 1.00 & 1.00 & 71.1 & 1.00 & 1.00 & 84.6 & 1.00 & 1.00 & 88.5 & 1.00 & 1.00 & 91.4 & 1.00 & 1.00 \\
 & EAGLE-3 & 88.0 & 5.93 & 5.93 & 84.9 & 5.35 & 5.35 & 71.3 & 3.71 & 3.71 & 84.2 & 3.71 & 3.71 & 87.4 & 3.90 & 3.90 & 91.3 & 3.85 & 3.85 \\
 & \fcRthree{SpecAttn} & \fcRthree{87.8} & \fcRthree{0.76} & \fcRthree{3.04} & \fcRthree{84.5} & \fcRthree{0.88} & \fcRthree{3.52} & \fcRthree{71.1} & \fcRthree{0.88} & \fcRthree{3.53} & \fcRthree{84.4} & \fcRthree{0.91} & \fcRthree{3.63} & \fcRthree{86.8} & \fcRthree{0.91} & \fcRthree{3.62} & \fcRthree{91.1} & \fcRthree{0.90} & \fcRthree{3.61} \\
 & \fcRthree{SpecPV} & \fcRthree{38.6} & \fcRthree{4.54} & \fcRthree{16.77} & \fcRthree{71.4} & \fcRthree{5.10} & \fcRthree{17.33} & \fcRthree{35.6} & \fcRthree{3.70} & \fcRthree{14.27} & \fcRthree{47.8} & \fcRthree{3.50} & \fcRthree{14.02} & \fcRthree{38.7} & \fcRthree{4.07} & \fcRthree{15.51} & \fcRthree{49.1} & \fcRthree{3.42} & \fcRthree{13.79} \\
 & \projName-lossy & 16.1 & 5.98 & 33.57 & 80.6 & 5.42 & 33.75 & 25.5 & 3.81 & 22.14 & 17.4 & 3.69 & 22.15 & 58.2 & 3.82 & 22.23 & 44.0 & 3.71 & 22.29 \\
 & \cellcolor{yellow!15}\projName-lossless & \cellcolor{yellow!15}87.8 & \cellcolor{yellow!15}3.52 & \cellcolor{yellow!15}10.53 & \cellcolor{yellow!15}85.0 & \cellcolor{yellow!15}3.27 & \cellcolor{yellow!15}10.43 & \cellcolor{yellow!15}71.3 & \cellcolor{yellow!15}2.10 & \cellcolor{yellow!15}7.23 & \cellcolor{yellow!15}84.4 & \cellcolor{yellow!15}1.99 & \cellcolor{yellow!15}7.02 & \cellcolor{yellow!15}88.3 & \cellcolor{yellow!15}2.48 & \cellcolor{yellow!15}8.55 & \cellcolor{yellow!15}91.3 & \cellcolor{yellow!15}2.43 & \cellcolor{yellow!15}8.30 \\
\midrule
\multirow{6}{*}{MATH500} & AR & 84.0 & 1.00 & 1.00 & 48.2 & 1.00 & 1.00 & 69.8 & 1.00 & 1.00 & 82.0 & 1.00 & 1.00 & 82.2 & 1.00 & 1.00 & 87.0 & 1.00 & 1.00 \\
 & EAGLE-3 & 85.0 & 5.23 & 5.23 & 48.8 & 4.16 & 4.16 & 67.6 & 4.24 & 4.24 & 81.8 & 3.73 & 3.73 & 80.4 & 3.78 & 3.78 & 86.6 & 3.91 & 3.91 \\
 & \fcRthree{SpecAttn} & \fcRthree{84.0} & \fcRthree{0.84} & \fcRthree{3.34} & \fcRthree{46.8} & \fcRthree{0.91} & \fcRthree{3.62} & \fcRthree{69.2} & \fcRthree{0.93} & \fcRthree{3.71} & \fcRthree{80.4} & \fcRthree{0.94} & \fcRthree{3.75} & \fcRthree{82.4} & \fcRthree{0.93} & \fcRthree{3.71} & \fcRthree{86.0} & \fcRthree{0.93} & \fcRthree{3.73} \\
 & \fcRthree{SpecPV} & \fcRthree{30.4} & \fcRthree{4.58} & \fcRthree{17.11} & \fcRthree{28.4} & \fcRthree{4.25} & \fcRthree{15.61} & \fcRthree{40.0} & \fcRthree{3.63} & \fcRthree{15.14} & \fcRthree{80.0} & \fcRthree{3.46} & \fcRthree{14.66} & \fcRthree{74.2} & \fcRthree{3.80} & \fcRthree{15.59} & \fcRthree{79.4} & \fcRthree{3.47} & \fcRthree{14.66} \\
 & \projName-lossy & 21.4 & 5.18 & 34.55 & 27.8 & 4.54 & 26.55 & 29.6 & 4.41 & 36.75 & 17.4 & 4.83 & 58.34 & 50.8 & 4.02 & 34.12 & 44.2 & 4.58 & 49.18 \\
 & \cellcolor{yellow!15}\projName-lossless & \cellcolor{yellow!15}84.0 & \cellcolor{yellow!15}3.80 & \cellcolor{yellow!15}11.51 & \cellcolor{yellow!15}48.6 & \cellcolor{yellow!15}2.83 & \cellcolor{yellow!15}9.38 & \cellcolor{yellow!15}69.0 & \cellcolor{yellow!15}2.89 & \cellcolor{yellow!15}9.68 & \cellcolor{yellow!15}82.0 & \cellcolor{yellow!15}2.10 & \cellcolor{yellow!15}7.07 & \cellcolor{yellow!15}81.2 & \cellcolor{yellow!15}2.89 & \cellcolor{yellow!15}10.02 & \cellcolor{yellow!15}87.0 & \cellcolor{yellow!15}2.76 & \cellcolor{yellow!15}9.55 \\
\midrule
\multirow{6}{*}{MBPP} & AR & 68.0 & 1.00 & 1.00 & 60.2 & 1.00 & 1.00 & 50.4 & 1.00 & 1.00 & 65.2 & 1.00 & 1.00 & 67.6 & 1.00 & 1.00 & 73.8 & 1.00 & 1.00 \\
 & EAGLE-3 & 68.0 & 4.96 & 4.96 & 59.8 & 4.37 & 4.37 & 51.4 & 3.97 & 3.97 & 65.4 & 3.44 & 3.44 & 65.8 & 3.12 & 3.12 & 73.8 & 3.30 & 3.30 \\
 & \fcRthree{SpecAttn} & \fcRthree{68.0} & \fcRthree{0.80} & \fcRthree{3.18} & \fcRthree{60.2} & \fcRthree{0.83} & \fcRthree{3.33} & \fcRthree{51.2} & \fcRthree{0.89} & \fcRthree{3.56} & \fcRthree{65.2} & \fcRthree{0.89} & \fcRthree{3.56} & \fcRthree{67.6} & \fcRthree{0.90} & \fcRthree{3.59} & \fcRthree{73.2} & \fcRthree{0.92} & \fcRthree{3.68} \\
 & \fcRthree{SpecPV} & \fcRthree{8.2} & \fcRthree{4.69} & \fcRthree{16.97} & \fcRthree{43.0} & \fcRthree{4.69} & \fcRthree{16.25} & \fcRthree{25.6} & \fcRthree{3.64} & \fcRthree{13.89} & \fcRthree{40.8} & \fcRthree{3.46} & \fcRthree{13.57} & \fcRthree{41.2} & \fcRthree{3.15} & \fcRthree{12.31} & \fcRthree{55.4} & \fcRthree{3.09} & \fcRthree{11.84} \\
 & \projName-lossy & 14.8 & 5.01 & 30.91 & 43.6 & 5.12 & 28.51 & 19.0 & 3.80 & 22.32 & 22.0 & 3.82 & 24.34 & 43.2 & 3.36 & 19.53 & 48.2 & 3.54 & 21.62 \\
 & \cellcolor{yellow!15}\projName-lossless & \cellcolor{yellow!15}68.0 & \cellcolor{yellow!15}2.74 & \cellcolor{yellow!15}8.59 & \cellcolor{yellow!15}60.2 & \cellcolor{yellow!15}2.82 & \cellcolor{yellow!15}8.66 & \cellcolor{yellow!15}50.6 & \cellcolor{yellow!15}2.05 & \cellcolor{yellow!15}6.86 & \cellcolor{yellow!15}65.4 & \cellcolor{yellow!15}1.97 & \cellcolor{yellow!15}6.60 & \cellcolor{yellow!15}66.2 & \cellcolor{yellow!15}2.52 & \cellcolor{yellow!15}8.14 & \cellcolor{yellow!15}73.8 & \cellcolor{yellow!15}2.53 & \cellcolor{yellow!15}8.46 \\
\bottomrule
\end{tabular}}
\vspace{-1em}
\end{table*}

\fcRthree{Fig.~\ref{fig:tapas_arch}(c) shows which ISU modules are active in the draft, lossy verification, and lossless verification phases, and how they overlap with MXU and SFU execution.}

\fcRone{The draft phase activates only the Accumulator.
As softmax results stream out of the Output SRAM, the Accumulator reads each attention weight from the write-out path, maps it to the corresponding chunk entry, and accumulates across all expansion steps of the round.
This tap adds no DRAM access beyond the baseline draft attention and does not stall the write-out path, and the draft phase therefore incurs no additional latency from the ISU.
At the end of the draft round, the Accumulator hands its chunk-level profile to the TopK Sorter, whose build and extract latency overlaps with subsequent token sampling and tree construction, keeping the Sorter off the end-to-end critical path.}

\fcRone{The lossy verification phase activates the KV Prefetcher, which translates the top-$K$ indices into DRAM addresses that drive selective KV cache loading.
While the MXU executes verify linear in W4A16, the Prefetcher issues $K$ chunk-sized burst reads to the Memory Controller.
Selective load volume is only $1/C_r$ of a full prefix load, and the linear-layer execution window is long enough to hide the prefetch latency.
By the time the MXU enters verify attention, the required KV entries are already resident in the Activation SRAM, and the number of KV entries participating in attention has dropped from $L_p$ to roughly $L_p / C_r$, yielding simultaneous savings in DRAM bandwidth and attention compute.}

\fcRone{The lossless verification phase disables all three ISU modules and reverts to standard tree-based SD verification.
The Activation SRAM loads the full prefix KV cache through standard ping-pong DMA.
KV rollback is realized by checkpointing the DRAM write pointer at the start of the lossy phase and restoring it when entering lossless verification, adding no DRAM access beyond the normal KV writes.}

%% file: chapters/5-evaluation.tex
\section{Evaluation} \label{sec:eval}

\subsection{Experimental Setup} \label{sec:eval_setup}

\textbf{Models, datasets, and baselines.}
We target LLM decoding in single-request \fcRtwo{mobile edge} scenarios with a batch size of 1.
We evaluate six target models spanning 1.7B to 14B parameters: DeepSeek-R1-Distill-Llama-8B (DeepSeek-R1-8B)~\cite{DeepSeek-R1}, Llama-3.1-8B-Instruct (Llama3.1-8B)~\cite{Llama3}, Qwen3-1.7B, Qwen3-4B, Qwen3-8B, and Qwen3-14B~\cite{Qwen3}.
We use three benchmarks: GSM8K~\cite{GSM8K} and MATH500~\cite{MATH500} for mathematical reasoning, and MBPP~\cite{MBPP} for code generation.
We compare \fcRthree{six} methods:
1)~AR~\cite{Transformer} is standard autoregressive decoding.
\fcRthree{2)~EAGLE-3~\cite{EAGLE-3} is the SotA tree-based SD method, with a draft tree of depth 5 and 60 nodes.}
\fcRthree{3)~SpecAttn~\cite{SpecAttn} is a lossless self-speculative method that drafts six tokens with the target model under sparse KV.}
\fcRthree{4)~SpecPV~\cite{SpecPV} is a lossy self-speculative method that shares the draft tree and $T_y$ of \projName, but refreshes the full KV without re-verifying the buffered chain.}
\fcRthree{5)}~\projName-lossy is our proposed sparse verification mode that uses draft attention scores to selectively load only the most important prefix KV chunks, skipping the rest to reduce DRAM traffic.
\fcRthree{6)}~\projName-lossless (\projName) is our proposed full two-pass scheme that alternates lossy and lossless verification to recover exact output quality while retaining most of the bandwidth savings.
All draft models are obtained from the EAGLE-3 public repository~\cite{EAGLE-3} with the default configurations.
For \projName, the compression ratio is set to $C_r = 8$, meaning each lossy step loads only $1/8$ of the prefix KV, and the chunk size is $C_s = 8$ to align with the hardware burst access granularity.
It additionally sets the lossy chain length via methods detailed in Sec.~\ref{sec:algorithm}.
\fcRthree{SpecAttn and SpecPV use the same KV budget of $C_r=8$ as \projName.}

\textbf{Algorithm-level setup.}
All algorithm experiments run on a single NVIDIA H100 GPU \fcRthree{in BF16 precision with greedy decoding} \fcRtwo{following the EAGLE-3 public repository~\cite{EAGLE-3}.}
Accuracy is evaluated with lm-evaluation-harness~\cite{lm_eval} under default configurations: GSM8K and MATH500 use chain-of-thought prompting with final-answer exact match, and MBPP uses pass@1 functional correctness.
\fcRtwo{With chain-of-thought prompting, a single reasoning trace spans thousands of tokens and pushes the context into the tens of thousands within one session.}
Two efficiency metrics are measured.
$\tau_\mathrm{step}$ is the average number of accepted tokens per decoding step.
$\tau_\mathrm{bw}$ is the bandwidth-normalized throughput defined in Sec.~\ref{sec:lossless_pass}, representing the number of tokens generated per full KV cache access.
For both metrics, higher values indicate better efficiency.
\fcRtwo{$\tau_\mathrm{step}$ and $\tau_\mathrm{bw}$ feed the system-level evaluation in Sec.~\ref{sec:eval_sys} as algorithm-level inputs.}


\textbf{Hardware-level setup.}
We integrate the ISU into the baseline edge NPU~\cite{Mokey, NPU-LLM} described in Sec.~\ref{sec:hardware}.
The NPU features an MXU with $64 \times 64$ PEs, 4\,MB on-chip SRAM, and 51.2\,GB/s off-chip DRAM bandwidth \fcRtwo{as shown in Table~\ref{tab:hw-spec}}.
Linear layers operate in W4A16 mode~\cite{GPTQ, AWQ, OmniQuant, Anda, FIGNA}, where weights are stored and transferred as INT4 and dequantized to BF16 inside the MXU for computation.
All other operations, including attention and nonlinear functions, are performed in full BF16 precision with BF16 data transferred from and written back to off-chip DRAM.
We implement the enhanced NPU in SystemVerilog and synthesize it with Synopsys Design Compiler under TSMC 16nm FinFET at 500\,MHz.
Power is obtained from Synopsys PrimeTime PX analysis with switching activity extracted from gate-level simulation on representative workloads.

\begin{figure}[t]
    \centering
    \includegraphics[width=\linewidth]{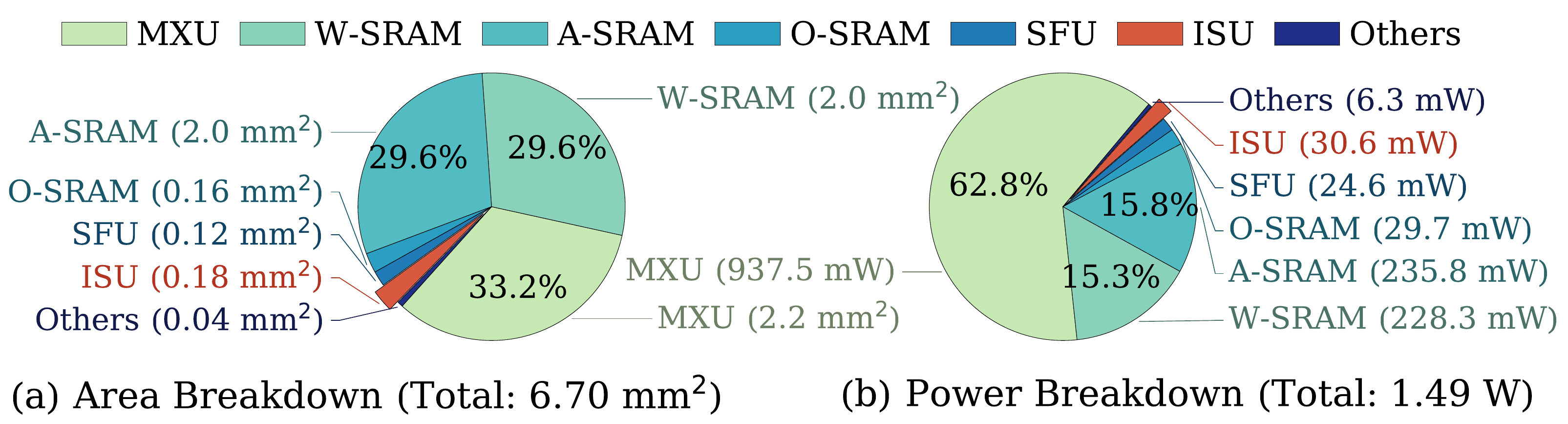}
    \vspace{-2em}
    \caption{Area and power breakdown of the \projName NPU. The ISU adds only 2.7\% area and 2.0\% power overhead.}
    \vspace{-1em}
    \label{fig:hw_overhead}
\end{figure}
\begin{figure}[t]
    \centering
    \includegraphics[width=\linewidth]{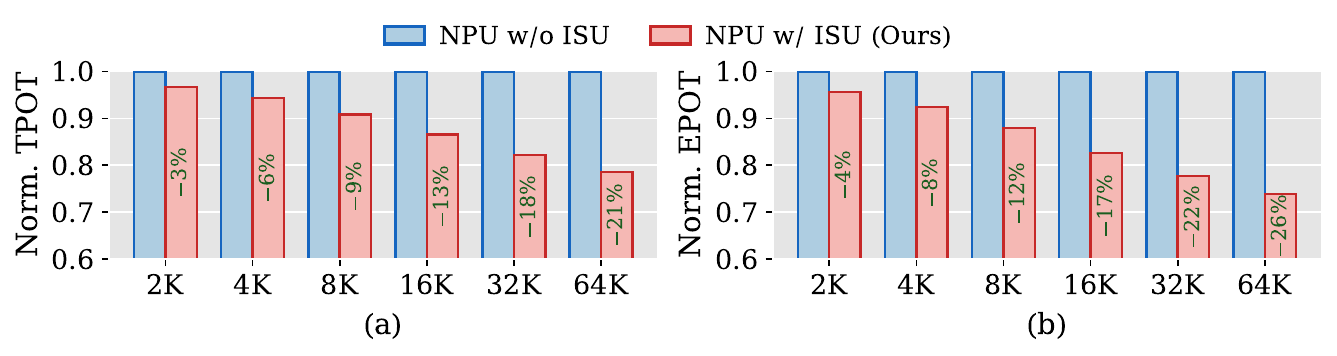}
    \vspace{-2em}
    \caption{\fcRtwo{Comparison of (a)~TPOT and (b)~EPOT w/ and w/o the ISU on Qwen3-8B. The ISU reduces TPOT by up to 21\% and EPOT by up to 26\%.}}
    \vspace{-1em}
    \label{fig:isu_ablation}
\end{figure}
\begin{figure*}[t]
    \centering
    \includegraphics[width=\linewidth]{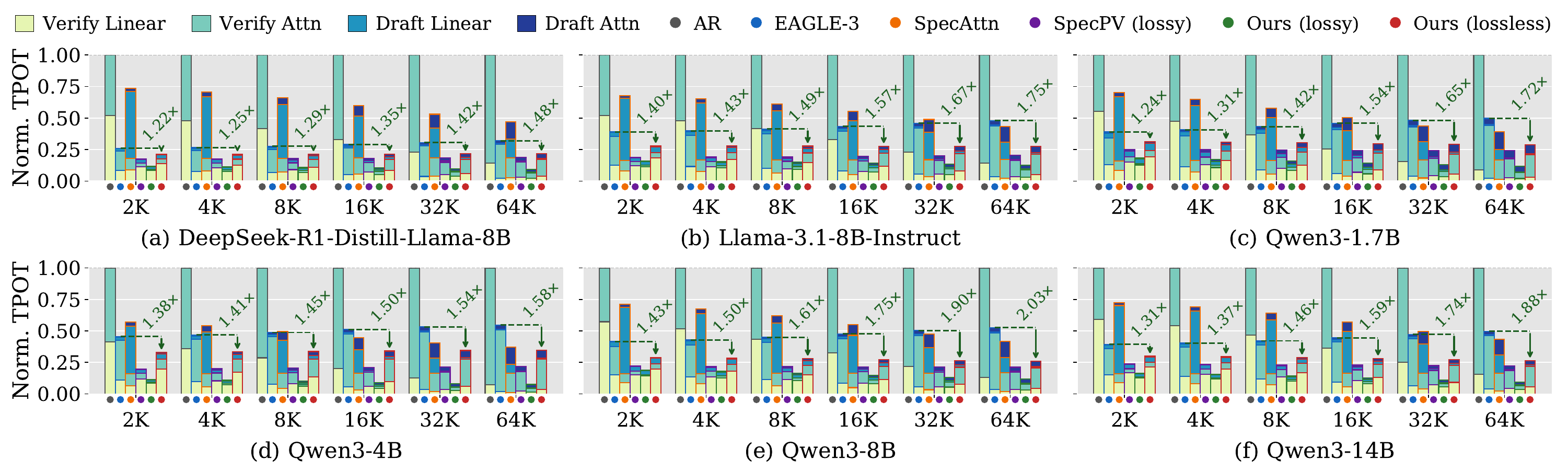}
    \vspace{-2em}
    \caption{\fcRthree{TPOT of EAGLE-3~\cite{EAGLE-3}, SpecAttn~\cite{SpecAttn}, SpecPV~\cite{SpecPV}, and \projName normalized to AR across six models and context lengths from 2K to 64K. Each bar is decomposed into verify linear, verify attention, draft linear, and draft attention. Lower bars indicate faster decoding.}}
    \vspace{-1em}
    \label{fig:sec5_speedup}
\end{figure*}

\textbf{System-level setup.}
\fcRtwo{Based on the \projName-enhanced NPU, we evaluate latency and energy using a cycle-level simulator following the evaluation methodology of prior NPU accelerator studies~\cite{BitMoD, Anda, FIGNA, Oaken}.}
\fcRtwo{The simulator instantiates each method's decoding workload from its algorithm-level evaluation in Sec.~\ref{sec:eval_algo}.}
\fcRthree{The simulator is cycle-accurate to our NPU implementation and further calibrated on an RTX 3090 GPU, with relative prediction errors within 8\% on the Jetson Thor and H200 GPUs of Fig.~\ref{fig:vllm}.}
\fcRthree{These measurements, $\tau_\mathrm{step}$ and $\tau_\mathrm{bw}$, are taken on MATH500 for each LLM and method, including SpecAttn and SpecPV.}
The default DRAM bandwidth is 51.2\,GB/s, corresponding to a four-channel LPDDR5 configuration~\cite{LPDDR5}, with an empirical efficiency factor of 0.7~\cite{Ramulator2}.
Off-chip DRAM energy is modeled at 6\,pJ/bit~\cite{DRAMsim3, Ramulator2, DRAMPower5}.
On-chip SRAM and compute energy are derived from synthesis power reports scaled by cycle count.
We report two system-level metrics: time-per-output-token (TPOT)~\cite{DistServe} and energy-per-output-token (EPOT)~\cite{Anda}, where lower values indicate better performance.

\subsection{Algorithm-level Evaluation} \label{sec:eval_algo}

This section evaluates \projName against AR~\cite{Transformer}, EAGLE-3~\cite{EAGLE-3}, \fcRthree{SpecAttn~\cite{SpecAttn}, and SpecPV~\cite{SpecPV}} on output quality and bandwidth efficiency.
We compare all methods across six models and three benchmarks in Table~\ref{tab:algo_main}.
\fcRthree{EAGLE-3, SpecAttn, and \projName-lossless} match AR accuracy on all combinations within GPU numerical noise, showing that \projName-lossless preserves the target model's output quality.
\fcRthree{In contrast, the lossy methods degrade unpredictably. \projName-lossy shows drops ranging from moderate (e.g., Llama3.1-8B GSM8K: 85.7\%\,$\to$\,80.6\%) to severe (e.g., Qwen3-4B GSM8K: 84.6\%\,$\to$\,17.4\%), and SpecPV, which commits its chain without re-verification, loses 2--60 points.}
For latency-sensitive scenarios that tolerate quality loss, \projName-lossy can be used directly, but for precision-sensitive tasks \projName-lossless is necessary. 

In terms of bandwidth efficiency, as shown in Table~\ref{tab:algo_main}, \projName-lossless achieves a lower $\tau_\mathrm{step}$ than EAGLE-3, approximately \fcRthree{20--50\%} lower across all combinations, reflecting the inherent cost of the two-pass mechanism where lossless re-verification rejects part of the lossy chain and adds extra steps.
Despite this, \projName-lossless achieves roughly $2\times$ higher $\tau_\mathrm{bw}$ than EAGLE-3.
Since EAGLE-3 loads the full prefix KV every verification step, its $\tau_\mathrm{bw}$ is identical to $\tau_\mathrm{step}$.
In contrast, most of \projName-lossless's decoding steps are lossy steps that load only $1/8$ of the prefix KV, and despite the increased step count the overall bandwidth efficiency remains substantially higher.
This higher $\tau_\mathrm{bw}$ means \projName-lossless generates more tokens per full KV cache access, an advantage that is particularly valuable on bandwidth-constrained mobile edge devices.
\fcRthree{SpecAttn reaches a $\tau_\mathrm{bw}$ of only 3.0--3.8, since every verification still loads the full KV cache, and its $\tau_\mathrm{step}$ stays below 1, since each sparse drafting pass also runs the target model.
SpecPV exceeds \projName-lossless in $\tau_\mathrm{bw}$ only by skipping re-verification, at the accuracy cost above, leaving \projName-lossless with the highest $\tau_\mathrm{bw}$ among lossless methods.}

\subsection{Hardware-level Evaluation} \label{sec:eval_hw}

This section quantifies the area and power overhead introduced by the ISU \fcRtwo{and the decoding gains it delivers in return}.
Fig.~\ref{fig:hw_overhead} shows the breakdown of the \projName NPU.
The total NPU occupies 6.70\,mm$^2$ and consumes 1.49\,W.
The MXU dominates both area and power, accounting for 33.2\% and 62.8\%, respectively.
Weight and Activation SRAMs together account for 59.1\% of area and 31.1\% of power.
The ISU adds only 0.18\,mm$^2$ of area and 30.6\,mW of power, accounting for just 2.7\% and 2.0\% of the NPU totals, respectively.
This low overhead arises because the ISU only performs lightweight importance profiling and top-K selection.

\fcRtwo{Fig.~\ref{fig:isu_ablation} isolates the benefit of the ISU on Qwen3-8B with MATH500, comparing our NPU against one that performs the same lossless \projName selection without dedicated hardware support.
The ISU lowers TPOT by 3\% at 2K and 21\% at 64K, and EPOT by 4\% and 26\%, respectively, with the gain widening as more KV chunks are scored and fetched at longer context.
The importance selection sits on the critical path, where scoring follows the draft attention results, the top-$K$ ranking follows the scoring of every expansion step, and each selected chunk then needs an explicit round trip before verify attention proceeds.
The ISU instead taps the Output SRAM write-out path, keeping the attention scores on chip and hiding their latency inside the token sampling and verify linear windows (Sec.~\ref{sec:hw_dataflow}), which removes the selection cost from both the critical path and the off-chip traffic.
The end-to-end gains enabled by this sub-3\% ISU hardware investment are evaluated next.}

\begin{figure*}[t]
    \centering
    \includegraphics[width=\linewidth]{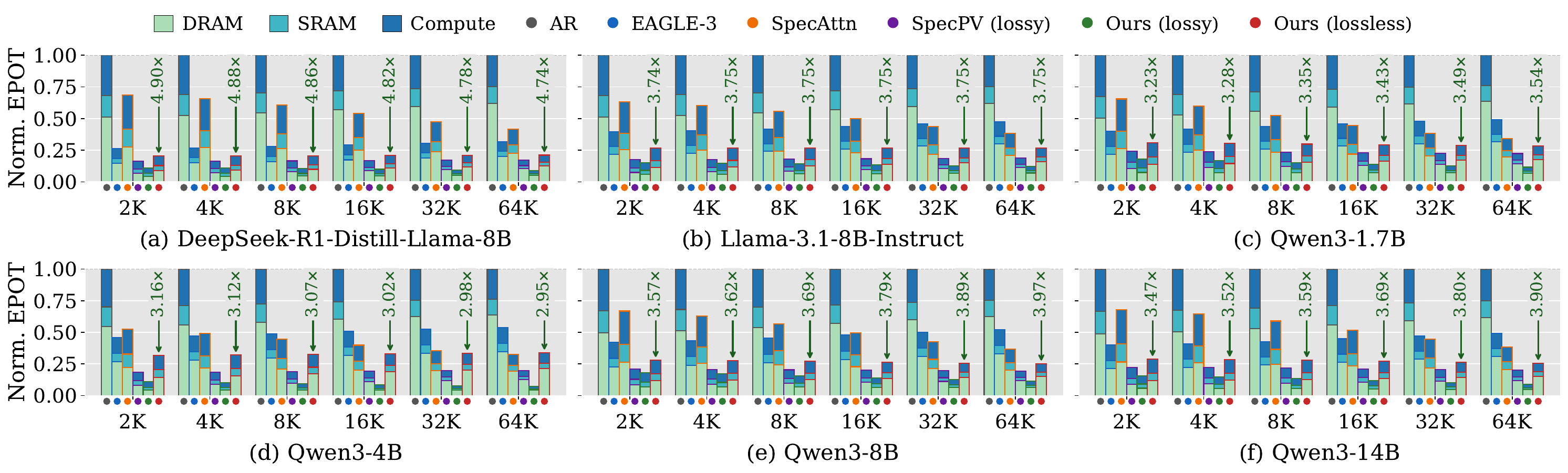}
    \vspace{-2em}
    \caption{\fcRthree{EPOT of EAGLE-3~\cite{EAGLE-3}, SpecAttn~\cite{SpecAttn}, SpecPV~\cite{SpecPV}, and \projName normalized to AR across six models and context lengths from 2K to 64K. Each bar is decomposed into DRAM, SRAM, and compute. Lower bars indicate better energy efficiency.}}
    \vspace{-1.5em}
    \label{fig:sec5_energy}
\end{figure*}

\subsection{System-level Evaluation} \label{sec:eval_sys}

This section evaluates how \projName's bandwidth savings translate into end-to-end latency and energy gains for edge deployment.
We measure two metrics that are critical for edge LLM decoding. TPOT directly determines the user-perceived decoding latency, and EPOT governs the per-token energy cost on resource-constrained edge devices.
We compare \projName against EAGLE-3~\cite{EAGLE-3}\fcRthree{, SpecAttn~\cite{SpecAttn},}\fcRtwo{ and SpecPV~\cite{SpecPV}} across six models and context lengths from \fcRtwo{2K} to \fcRtwo{64K}, covering scenarios from short-form chat to long-context agentic workloads.

Fig.~\ref{fig:sec5_speedup} shows the TPOT \fcRthree{normalized to AR} across all configurations.
On average, \projName-lossless achieves \fcRtwo{1.33$\times$ to 1.74$\times$} lower TPOT than EAGLE-3, with the highest gain of \fcRtwo{2.03$\times$} on Qwen3-8B \fcRtwo{at 64K}.
\fcRtwo{The advantage widens steadily with context, from 1.33$\times$ at 2K to 1.74$\times$ at 64K, as the growing prefix KV makes verification increasingly bandwidth-bound.}
Moreover, the per-category breakdown in Fig.~\ref{fig:sec5_speedup} reveals the underlying trade-off.
Compared to EAGLE-3, \projName-lossless dramatically reduces verify attention thanks to higher $\tau_\mathrm{bw}$ from sparse KV loading, while verify linear and draft operations occupy a relatively larger share due to the lower $\tau_\mathrm{step}$ introduced by two-pass decoding.
For example, on Qwen3-8B at \fcRtwo{64K}, EAGLE-3's verify attention accounts for \fcRtwo{86\%} of its total cost, whereas \projName-lossless reduces it to \fcRtwo{62\%}, yielding an overall \fcRtwo{2.03$\times$} TPOT improvement.
\fcRthree{SpecAttn reloads the target weights for every drafted token, which makes it slower than EAGLE-3 at short context, and \projName is 1.7--3.5$\times$ faster than it at 2K and 1.1--2.2$\times$ at 64K.
At the same KV budget, \projName-lossy is 1.2--2.9$\times$ faster than SpecPV, which reloads the full KV at every refresh.
SpecPV outpaces \projName-lossless only by skipping re-verification, at the accuracy cost in Table~\ref{tab:algo_main}.}
\fcRtwo{Across both comparisons, \projName's advantage grows with context, and it continues to widen as reasoning-driven agentic workloads push context lengths even further.}


Fig.~\ref{fig:sec5_energy} shows the EPOT \fcRthree{normalized to AR} across the same configurations.
\fcRtwo{On average, \projName-lossless achieves 1.40$\times$ to 1.76$\times$ lower EPOT than EAGLE-3, with the highest gain of 2.06$\times$ on Qwen3-8B at 64K.}
Edge NPU energy consumption consists of three components: off-chip DRAM access, on-chip SRAM access, and compute.
\fcRtwo{The energy breakdown in Fig.~\ref{fig:sec5_energy} shows that \projName-lossless reduces the DRAM term that dominates edge decoding energy, while its compute and SRAM shares grow modestly under the lower $\tau_\mathrm{step}$ of two-pass decoding.}
\fcRtwo{The EPOT gain stays ahead of the TPOT gain at every context length,
since DRAM energy is proportional to total access volume and every reduction in KV traffic translates directly into energy savings.}
\fcRthree{The same ordering holds in energy, where \projName-lossy is 1.2--2.8$\times$ more efficient than SpecPV and \projName-lossless up to 3.4$\times$ more than SpecAttn, with parity only on Qwen3-4B at 64K.}
This makes \projName particularly attractive for energy-constrained edge devices, where reducing DRAM access is one of the most effective levers for extending battery life.

\begin{figure}[t]
    \centering
    \includegraphics[width=\linewidth]{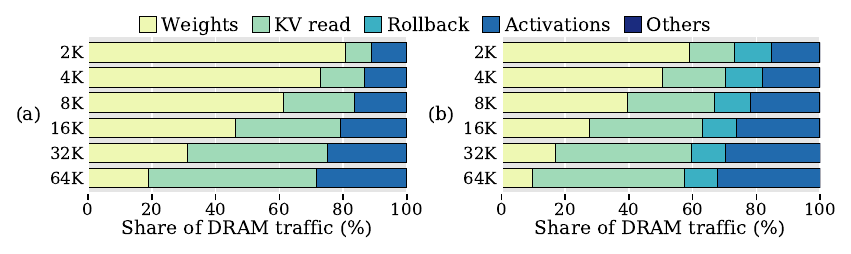}
    \vspace{-2em}
    \caption{\fcRthree{DRAM traffic of (a)~lossy and (b)~lossless \projName on Qwen3-8B.}}
    \vspace{-1em}
    \label{fig:dram_traffic}
\end{figure}

\fcRthree{Fig.~\ref{fig:dram_traffic} further breaks down the DRAM traffic per output token of \projName on Qwen3-8B, whose classes sum to the traffic behind Figs.~\ref{fig:sec5_speedup} and~\ref{fig:sec5_energy}.
Weights dominate at short context and KV reads at long context, while the others, KV writes and selection metadata, stay below 0.2\%.
Rollback, the traffic of lossy steps later rejected, takes a stable 10--12\% of the traffic of \projName-lossless and is already included in its TPOT and EPOT.
\projName-lossless roughly halves EAGLE-3's per-token DRAM traffic, saving ten times its rollback cost.}

\subsection{Discussion} \label{sec:eval_disc}

\fcRthree{Beyond one hardware configuration and one set of algorithm parameters, we next examine how DRAM bandwidth, acceptance rate, batch size and numeric format, model scale and workload, and batching on GPUs affect \projName's gains.}

\begin{figure}[t]
    \centering
    \includegraphics[width=\linewidth]{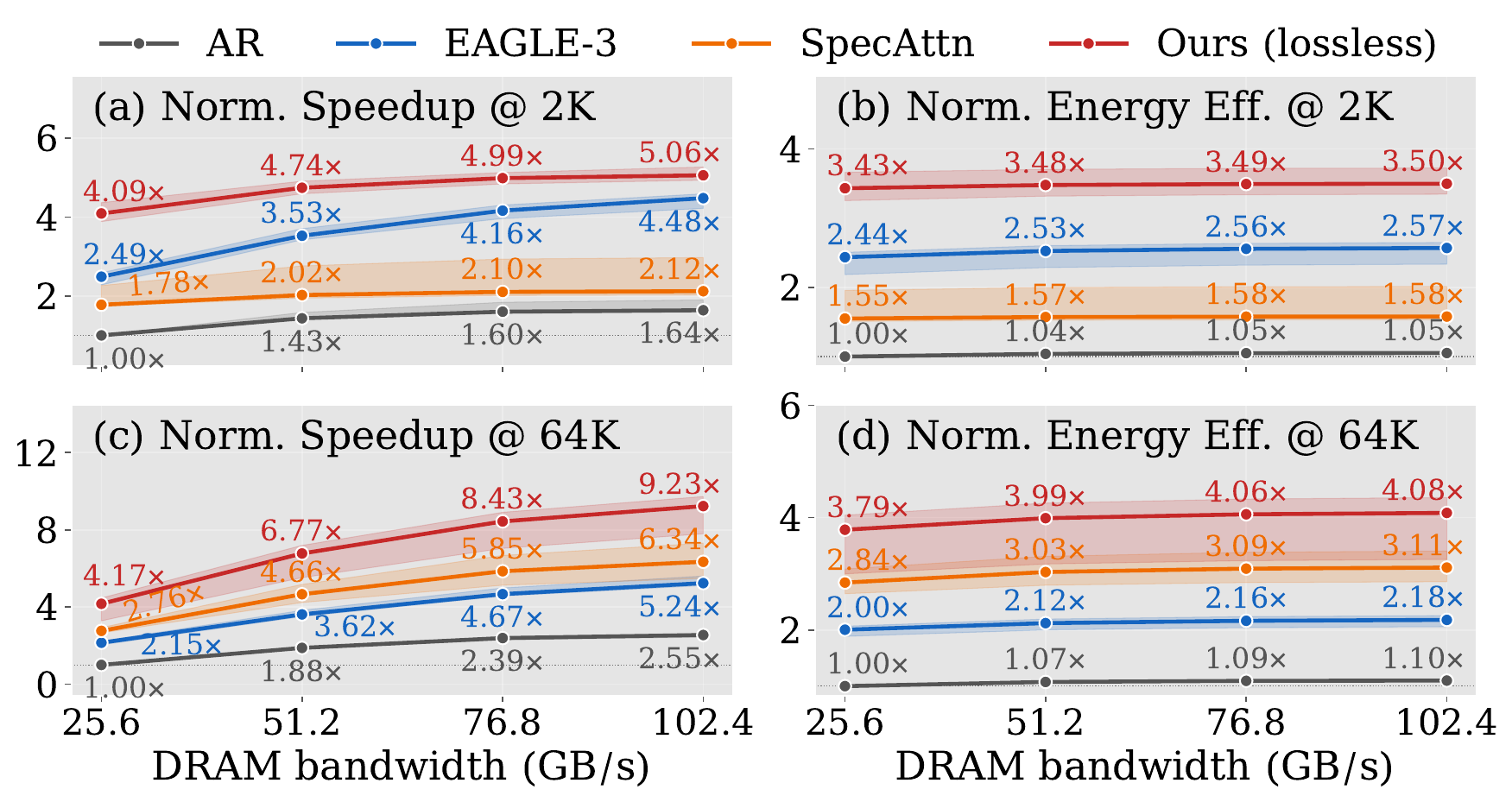}
    \vspace{-2em}
    \caption{\fcRthree{Normalized speedup and energy efficiency of AR, EAGLE-3, SpecAttn, and \projName vs.\ DRAM bandwidth at 2K and 64K context. Each line shows the median of four Qwen3 models, normalized to AR at 25.6\,GB/s.}}
    \vspace{-1.5em}
    \label{fig:bw_dse}
\end{figure}

\textbf{Impact of DRAM bandwidth.}
Mobile edge devices employ a variety of DRAM technologies, with bandwidths ranging from around 25\,GB/s for LPDDR4~\cite{LPDDR4} to over 100\,GB/s for LPDDR5X~\cite{LPDDR5X} \fcRtwo{as shown in Table~\ref{tab:hw-spec}}.
To assess \projName's robustness across this \fcRtwo{range}, Fig.~\ref{fig:bw_dse} sweeps the bandwidth from 25.6 to 102.4\,GB/s while keeping all other hardware parameters fixed.
\projName is compared with AR~\cite{Transformer}, EAGLE-3~\cite{EAGLE-3}, \fcRthree{and SpecAttn~\cite{SpecAttn}} using the Qwen3 family (1.7B/4B/8B/14B)~\cite{Qwen3} at 2K and \fcRtwo{64K} context lengths. We report normalized speedup and energy efficiency, both relative to AR at the lowest bandwidth of 25.6\,GB/s, where higher values indicate better performance.
All methods benefit from higher bandwidth in terms of speedup, but the relative advantage of \projName over EAGLE-3 is most pronounced at low bandwidth, where the verification bandwidth bottleneck is most severe. At \fcRtwo{64K} context, \projName achieves \fcRtwo{4.17$\times$} speedup at 25.6\,GB/s and \fcRtwo{9.23$\times$} at 102.4\,GB/s, corresponding to \fcRtwo{1.94$\times$} and \fcRtwo{1.76$\times$} gains over EAGLE-3, respectively. At 2K context the trend is consistent, with \projName achieving \fcRtwo{4.09$\times$} to 5.06$\times$ speedup across the bandwidth range.
In contrast, energy efficiency is nearly insensitive to bandwidth, with \projName at \fcRtwo{64K} context ranging from \fcRtwo{3.79$\times$} to \fcRtwo{4.08$\times$}, mainly because the energy saved by reducing DRAM accesses depends on per-bit energy cost rather than transfer speed.
\fcRthree{SpecAttn remains below \projName across the whole bandwidth range.}
Overall, \projName delivers consistent gains over EAGLE-3 across the bandwidth range, with the speedup advantage growing on more bandwidth-constrained platforms~\cite{MobileSoC, AppleIntelligence, SamsungGalaxyAI}.

\begin{figure}[t]
    \centering
    \includegraphics[width=\linewidth]{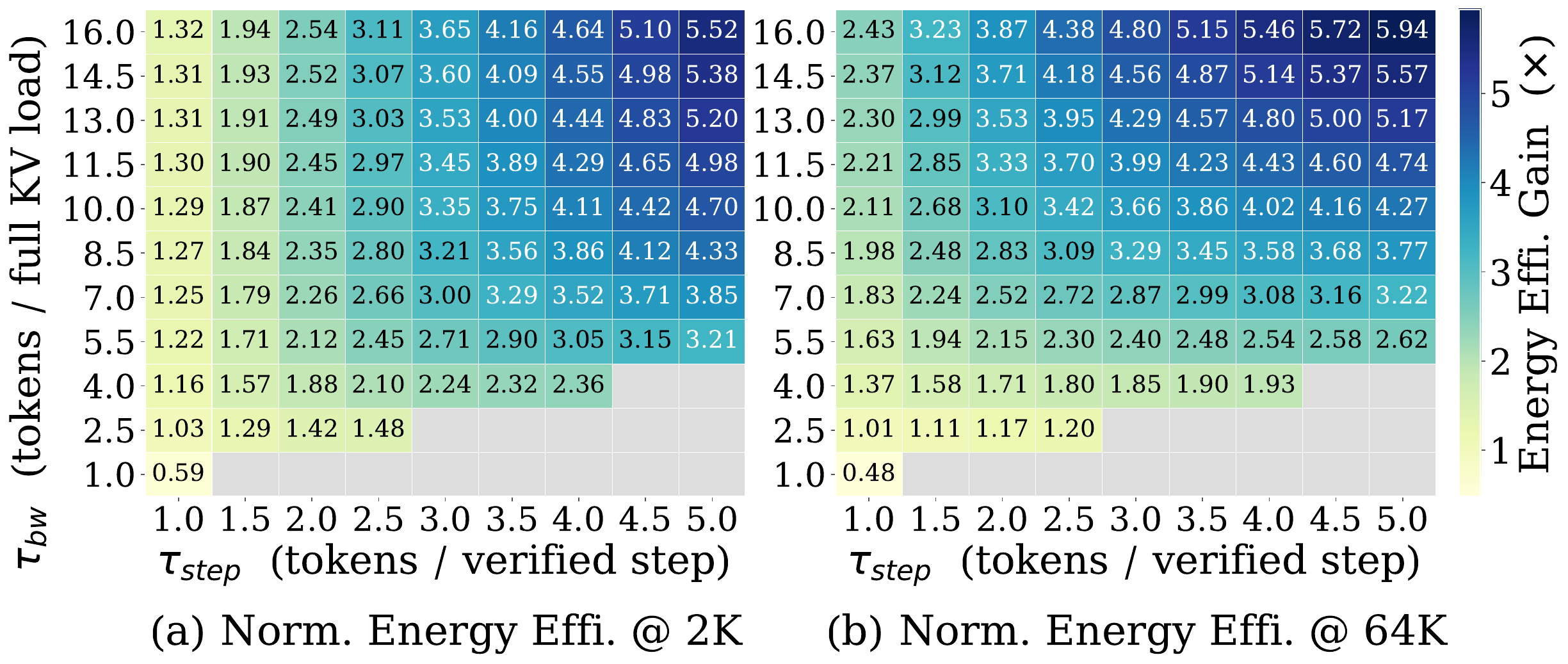}
    \vspace{-2em}
    \caption{\fcRtwo{Energy efficiency gain over AR across varying $\tau_\mathrm{step}$ and $\tau_\mathrm{bw}$ at 2K and 64K context lengths on Qwen3-1.7B with 51.2\,GB/s DRAM bandwidth.}}
    \vspace{-1em}
    \label{fig:tau_dse}
\end{figure}

\textbf{Impact of acceptance rate.}
To understand how future improvements to \projName's acceptance rate would translate into hardware-level gains, we sweep $\tau_\mathrm{step}$ and $\tau_\mathrm{bw}$ independently across a wide range of values in our system-level simulation.
We report energy efficiency normalized to AR as the metric.
Taking Qwen3-1.7B as an example, Fig.~\ref{fig:tau_dse} maps the energy efficiency gain over AR at 2K and \fcRtwo{64K} context lengths, representing short-form and long-context scenarios, respectively.
At 2K context, the heatmap gradient is steeper along $\tau_\mathrm{step}$, indicating that improving acceptance rate yields greater marginal benefit when compute dominates.
At \fcRtwo{64K} context, the gradient shifts to $\tau_\mathrm{bw}$, indicating that reducing per-step KV access volume matters more when KV loading dominates total energy.
We further examine the $\geq$3$\times$ efficiency region as a practical deployment target.
At 2K context, reaching 3$\times$ requires $\tau_\mathrm{step}$\,=\,3.0 and $\tau_\mathrm{bw}$\,=\,10.0 to achieve 3.35$\times$. \fcRtwo{At 64K context, the $\tau_\mathrm{step}$ threshold drops to 2.0 at the same $\tau_\mathrm{bw}$\,=\,10.0, already reaching 3.10$\times$.}
This shows that in long-context scenarios, even a modest acceptance rate can deliver over 3$\times$ energy efficiency gains provided the bandwidth savings are sufficient.
As context lengths continue to grow with \fcRtwo{emerging reasoning-driven agentic workloads}~\cite{MobileSoC, FastTTS, MobileAgent, AppAgent, EdgeReasoning}, improving $\tau_\mathrm{bw}$ in future algorithm development can effectively translate into better hardware-level decoding efficiency.

\begin{figure}[t]
    \centering
    \includegraphics[width=\linewidth]{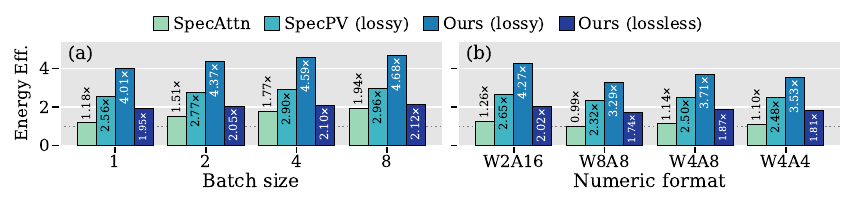}
    \vspace{-2.2em}
    \caption{\fcRthree{Energy efficiency over EAGLE-3 on Qwen3-8B at 32K context across (a)~batch sizes and (b)~numeric formats.}}
    \vspace{-1.8em}
    \label{fig:batch_precision}
\end{figure}

\fcRthree{\textbf{Impact of batch size and numeric format.}
Fig.~\ref{fig:batch_precision} sweeps the batch size and the numeric format on Qwen3-8B at 32K context.
Each format quantizes the linear activations and the KV cache to the same bit width, while the MAC array remains in BF16 and $\tau$ is taken from BF16 runs.
Batching amortizes the weight traffic across sequences, whereas each sequence still reads its own KV cache.
\projName's gain over EAGLE-3 therefore grows from 1.95$\times$ to 2.12$\times$ as the batch size increases from 1 to 8.
KV quantization reduces the bytes per KV entry, whereas \projName reduces the number of entries loaded, and the two compose.
Even under W4A4, \projName still halves EAGLE-3's per-token DRAM traffic, and its gain drops only from 1.95$\times$ under W4A16 to 1.87$\times$ under W4A8 and 1.81$\times$ under W4A4.}

\begin{table}[t]
\centering
\caption{\fcRthree{Long-context evaluation on LongBench.}}
\label{tab:algo_longbench}
\resizebox{\columnwidth}{!}{%
\begin{tabular}{lccc|ccc}
\toprule
 & \multicolumn{3}{c|}{\fcRthree{\textbf{Qwen3-8B (GovReport / QMSum)}}} & \multicolumn{3}{c}{\fcRthree{\textbf{Qwen3-32B (GovReport / QMSum)}}} \\
\cmidrule(lr){2-4} \cmidrule(lr){5-7}
\fcRthree{\textbf{Method}} & \fcRthree{ROUGE} & \fcRthree{$\tau_\mathrm{step}$} & \fcRthree{$\tau_{bw}$} & \fcRthree{ROUGE} & \fcRthree{$\tau_\mathrm{step}$} & \fcRthree{$\tau_{bw}$} \\
\midrule
\fcRthree{EAGLE-3} & \fcRthree{34.2~/~24.8} & \fcRthree{3.30~/~2.88} & \fcRthree{\phantom{0}3.30~/~2.88\phantom{0}} & \fcRthree{33.5~/~24.2} & \fcRthree{2.55~/~2.31} & \fcRthree{\phantom{0}2.55~/~2.31\phantom{0}} \\
\fcRthree{SpecAttn} & \fcRthree{34.6~/~24.7} & \fcRthree{0.87~/~0.91} & \fcRthree{\phantom{0}3.48~/~3.64\phantom{0}} & \fcRthree{33.5~/~23.6} & \fcRthree{0.87~/~0.90} & \fcRthree{\phantom{0}3.49~/~3.61\phantom{0}} \\
\fcRthree{SpecPV} & \fcRthree{33.2~/~23.8} & \fcRthree{3.20~/~2.58} & \fcRthree{10.70~/~8.90\phantom{0}} & \fcRthree{32.6~/~24.8} & \fcRthree{2.48~/~2.23} & \fcRthree{\phantom{0}9.18~/~8.34\phantom{0}} \\
\fcRthree{\projName-lossy} & \fcRthree{33.2~/~24.5} & \fcRthree{3.31~/~2.88} & \fcRthree{25.56~/~22.28} & \fcRthree{32.8~/~24.3} & \fcRthree{2.58~/~2.31} & \fcRthree{19.84~/~17.89} \\
\rowcolor{yellow!15}\fcRthree{\projName-lossless} & \fcRthree{34.2~/~25.0} & \fcRthree{1.77~/~1.69} & \fcRthree{\phantom{0}5.95~/~5.82\phantom{0}} & \fcRthree{33.2~/~23.4} & \fcRthree{1.23~/~1.02} & \fcRthree{\phantom{0}4.55~/~3.84\phantom{0}} \\
\bottomrule
\end{tabular}}
\vspace{-1em}
\end{table}

\begin{figure}[t]
    \centering
    \includegraphics[width=\linewidth]{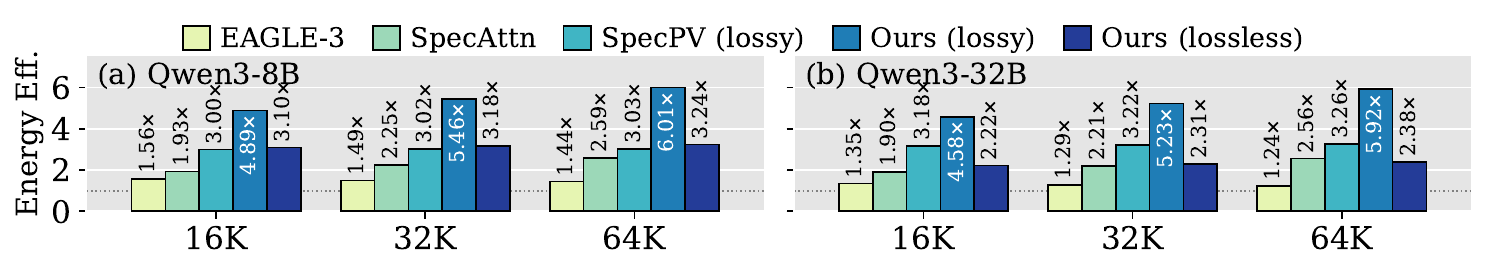}
    \vspace{-2em}
    \caption{\fcRthree{Energy efficiency over AR on QMSum from LongBench.}}
    \vspace{-1em}
    \label{fig:longbench_epot}
\end{figure}

\fcRthree{\textbf{Impact of model scale and long-context workloads.}
Table \ref{tab:algo_longbench} extends the evaluation to Qwen3-32B~\cite{Qwen3} and two LongBench~\cite{LongBench} tasks, GovReport and QMSum, with 50 documents per task and prompts of 3K to 32K tokens.
All methods reach comparable ROUGE-L.
\projName-lossless achieves 1.7--2.0$\times$ the $\tau_\mathrm{bw}$ of EAGLE-3 on both models, in line with Table~\ref{tab:algo_main}, and 1.1--1.7$\times$ that of SpecAttn, with the gap narrowing on Qwen3-32B.
As Fig.~\ref{fig:longbench_epot} shows for QMSum, this yields 2.0--2.3$\times$ lower EPOT than EAGLE-3 on Qwen3-8B and 1.6--1.9$\times$ on Qwen3-32B, growing from 16K to 64K.
\projName's advantage thus persists for a 32B model and long documents.}

\begin{figure}[t]
    \centering
    \includegraphics[width=\linewidth]{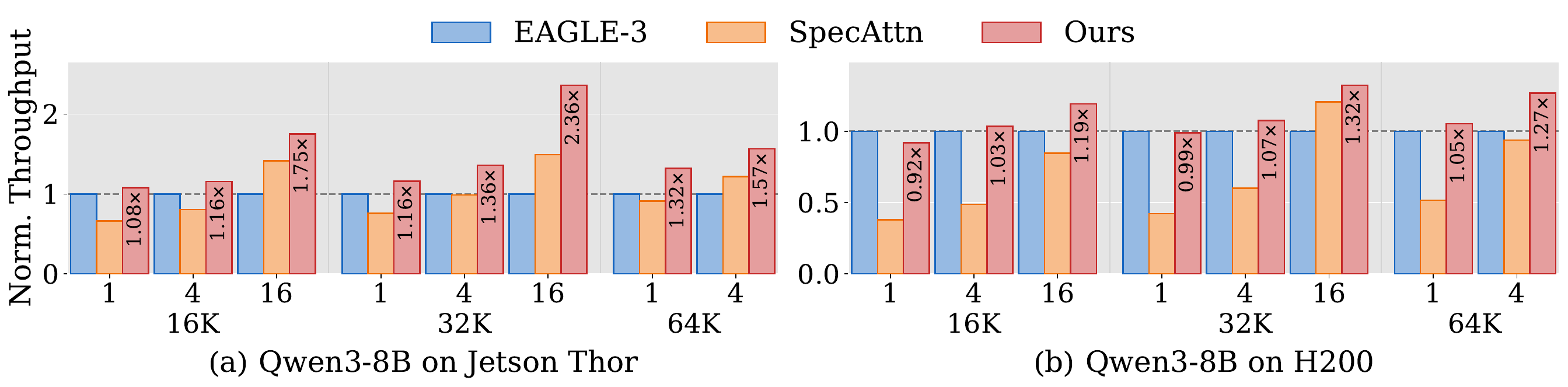}
    \vspace{-2em}
    \caption{\fcRthree{Throughput normalized to EAGLE-3 on Qwen3-8B on (a)~Jetson Thor and (b)~H200, where batch size 16 at 64K runs out of memory.}}
    \vspace{-1em}
    \label{fig:vllm}
\end{figure}

\phantomsection\label{par:vllm}
\fcRthree{\textbf{Impact of batching on GPUs.}
We further implement \projName in vLLM~\cite{vLLM} in full BF16 precision on a Jetson Thor edge GPU and an H200 cloud GPU, and decode Qwen3-8B on LongBench-v2~\cite{LongBenchV2} inputs of 16K--64K tokens.
Under greedy decoding, its outputs are identical to those of EAGLE-3 with the same verification length.
As Fig.~\ref{fig:vllm} shows, \projName is the fastest in every setting, except at batch size 1 with 16K and 32K context on the H200, where it stays within 8\% of EAGLE-3 and remains far ahead of SpecAttn.
On Thor, whose bandwidth is far below that of the H200, KV loading dominates, and \projName outpaces EAGLE-3 in every setting, including batch size 1.
On both GPUs, its gain grows with context and batch size, reaching 2.36$\times$ on Thor and 1.32$\times$ on the H200, tracking KV traffic rather than the platform.}

%% file: chapters/6-related.tex
\section{Related Work} \label{sec:related}


\fcRtwo{\textbf{Lossy decoding.}}
\fcRone{Lossy methods cut KV traffic by giving up exact fidelity.
\fcRtwo{\emph{Quantization}~\cite{GPTQ, AWQ, OmniQuant, TurboQuant, Oaken, COMET, FIGNA, BitMoD, Anda, Mokey, APT-LLM} shrinks the bytes per token.}
It still loads every token, and being orthogonal to \projName it stacks on top of our lossless path.
\fcRtwo{\emph{KV cache sparsification}~\cite{H2O, SnapKV, SCOPE, SAGE-KV, SkipKV, StreamingLLM, AttentionSink, KVCompression, SpeculativePrefill, HiKV} instead reduces the number of tokens loaded, keeping only a subset and evicting the rest.}
\fcRtwo{Sparsification saves the most bandwidth but evicts irreversibly, and a token that turns important later can no longer be recovered.
\projName-lossy loads the same reduced set while retaining the full KV cache, using draft attention only to steer which tokens are loaded.}}

\fcRtwo{\textbf{Lossless decoding.}}
\fcRone{\fcRthree{Lossless methods amortize KV loading over more generated tokens while preserving the target model's output distribution.}
\emph{Speculative decoding}~\cite{SD, Medusa, EAGLE, EAGLE-2, EAGLE-3, Sequoia, Hydra, SpecInfer, CSD, SpecExtend, LongSpec, DFlash, MTP, DSpark} pairs a lightweight draft with a target verifier and accepts many tokens per step, spreading one KV load across them.
\fcRthree{Tree-based variants~\cite{EAGLE, EAGLE-2, EAGLE-3, HASS, GTO, DDTree} verify many candidate paths in one forward pass.}
\emph{Self-speculative decoding}~\cite{DraftVerify, LayerSkip, Kangaroo, SWIFT, SpecEE, SpecAttn, SpecPV, Cassandra} drafts with the target model itself through early exiting or layer skipping, saving the draft weight traffic rather than the KV traffic.
\fcRthree{\projName is faster than the self-speculative SpecAttn and more accurate than SpecPV, and suits any lightweight draft.
As Fig.~\ref{fig:spd_profile} shows, verification dominates for drafts up to the five-layer block-diffusion drafter of DDTree~\cite{DDTree}, and multi-token prediction heads~\cite{MTP} are lighter still.}
\fcRtwo{Amortization lowers the load per generated token, yet each verification step still reads entire prefix KV.}}
\fcRtwo{SpecPV~\cite{SpecPV} reduces this load, yet re-scores candidate chunks at every step, trading prefix traffic for scoring traffic.
\projName instead reuses the scores the draft model has already produced.}

\fcRtwo{\textbf{Algorithm-hardware co-design.}
LLM accelerators raise effective bandwidth or shrink weight and KV traffic through low-precision datatypes~\cite{FIGNA, BitMoD, Anda, Mokey, APT-LLM, M2XFP, M-ANT, OliVe} or near-memory computing~\cite{NeuPIMs, AttAcc, AttenPIM, UM-PIM, P3-LLM}, both of which redesign the datapath or the memory hierarchy.
Hardware-level speculative decoding couples the draft-verify loop with early exiting, sparse attention, or edge-cloud partitioning~\cite{SpecEE, DSD, EdgeLLM, SpecAttn, MagicDec, SpecInfer, SpecIllusion, LongSpec}, leaving the verification KV bandwidth untouched.
Rather than reworking the datapath or the memory hierarchy, \projName shows that a minimal extension to a conventional NPU~\cite{SamsungNPU, NPU-LLM, MobileSoC, tpu, tpuv4i, OliVe, Tender} is enough, where a tiny ISU turns draft attention into the bandwidth signal for verification, unlocking the efficient bandwidth utilization.}
 

%% file: chapters/7-conclusion.tex
\section{Conclusion} \label{sec:conclusion}


\fcRone{This work presents \projName, an algorithm-hardware co-design that tackles the verification bandwidth bottleneck of tree-based speculative decoding on \fcRtwo{mobile edge} devices.}
\fcRone{Our key insight is that draft attention scores, readily available before verification, serve as a reliable proxy for prefix importance across diverse models and tasks.}
\fcRone{On the algorithm side, \projName combines a lossy sparse-verification pass guided by draft attention with a periodic lossless re-verification pass, amortizing its bandwidth cost over many accepted tokens.}
\fcRone{On the hardware side, \projName extends a conventional edge NPU with a lightweight Importance Selection Unit.}
\fcRone{Evaluated across six LLMs and three benchmarks against the state-of-the-art EAGLE-3, \projName achieves \fcRtwo{1.33--1.74$\times$} speedup and \fcRtwo{1.40--1.76$\times$} energy efficiency \fcRtwo{across 2K--64K contexts}, with only 2.7\% area overhead over the baseline edge NPU.}
\fcRone{As \fcRtwo{reasoning-driven} agentic workloads push context lengths ever higher, \projName demonstrates a scalable path toward efficient edge LLM decoding.}

%% file: chapters/10-AI_use.tex
\section*{AI Use}


We used Claude for language polishing and coding assistance. The authors reviewed and take responsibility for the AI-assisted text and code. The research ideas, experimental design, results, and conclusions are the authors' own.